\documentclass[preprint,3p,12pt]{elsarticle}
\usepackage{graphicx} % Required for inserting images
\usepackage{subfigure}
\usepackage{amsmath,amsfonts,amssymb}
\usepackage{algorithm}
\usepackage{algorithmic}

\newcommand{\Rmnum}[1]{\uppercase\expandafter{\romannumeral #1}}
\begin{document}
\title{Adaptive finite volume-particle method for free surface flows}

\author[HKUST1]{Jiawang Zhang}
\ead{jzhangiw@connect.ust.hk}

\author[HKUST1]{Fengxiang Zhao}
\ead{fzhaoac@connect.ust.hk}

\author[HKUST1,HKUST2,HKUST3]{Kun Xu\corref{cor1}}
\ead{makxu@ust.hk}

\address[HKUST1]{Department of Mathematics, Hong Kong University of Science and Technology, Clear Water Bay, Kowloon, Hong Kong}
\address[HKUST2]{Center for Ocean Research in Hong Kong and Macau (CORE), Hong Kong University of Science and Technology, Clear Water Bay, Kowloon, Hong Kong}
\address[HKUST3]{Shenzhen Research Institute, Hong Kong University of Science and Technology, Shenzhen, China}
\cortext[cor1]{Corresponding author}

\begin{abstract}

    This study proposes a novel adaptive finite volume-particle method (AFVPM) for accurate and efficient free surface flow simulations. The proposed AFVPM synergistically combines the Eulerian finite volume method (FVM) on unstructured meshes with the Lagrangian smoothed particle hydrodynamics (SPH) approach. Specifically, the mesh-based FVM is employed in the bulk flow regions to leverage its computational efficiency and numerical accuracy, while a weakly compressible SPH formulation is applied in the vicinity of the interface to maintain robust free-surface tracking capabilities. A key innovation of this framework is a block-based dynamic and adaptive conversion strategy between Eulerian mesh regions and Lagrangian particle regions and a buffer region-based cell-particle algorithm is designed to ensure seamless data communication across the Eulerian mesh–Lagrangian particle interface. Furthermore, isothermal gas-kinetic scheme (GKS) incorporating gravitational effects is utilized to calculate the fluxes in the mesh regions. The performance and reliability of the proposed AFVPM are validated through a series of benchmark cases that involve complex free surface phenomena. Numerical results demonstrate that AFVPM achieves superior accuracy and efficiency compared to full SPH approaches.
\end{abstract}
\begin{keyword}
    Smoothed particle hydrodynamics, Finite volume method, Gas-kinetic scheme, Free surface flow
\end{keyword}

\maketitle

\section{Introduction}
Free-surface flows, which involve either a free liquid surface or interfaces separating immiscible fluids, are ubiquitous and critically important in numerous natural phenomena \cite{waveBreak} and  engineering applications, attracting widespread attention from both the scientific and industrial communities \cite{multiphase}. However, the complex dynamics of moving interfaces, including breaking, fragmentation, and reconnection, continue to pose significant challenges for the efficient and accurate simulation of free-surface flows.

Most existing numerical methods for free-surface flows are based on mesh-based frameworks combined with interface-capturing or front-tracking techniques, such as the Volume-of-Fluid (VOF) method \cite{vof}, the Level Set (LS) method \cite{ls}, and their variants \cite{clsvof}. In mesh-based methods, the computational domain is discretized into a set of cells on which the governing equations are solved, and the flow field is represented by cell-averaged or nodal values. These methods are widely recognized for their high computational efficiency and accuracy, supported by well-established theories of stability, consistency, and convergence. However, accurately reconstructing the location of free surfaces on a fixed Eulerian mesh is often complicated and computationally expensive \cite{thinc,2ndvof}. Moreover, mesh-based methods inevitably introduce numerical diffusion near the free surface, which tends to smear the interface.

In view of these limitations, mesh-free particle-based methods have attracted increasing attention for free-surface flow simulations, with Smoothed Particle Hydrodynamics (SPH) being a representative example \cite{sph1,sph2}. In SPH, the computational domain is represented by a set of arbitrarily distributed particles, and the evolution of field variables associated with each particle, such as density and velocity, is obtained through integral approximations of the Lagrangian governing equations, evaluated as summations over neighboring particles \cite{liusph}. As the particles move, the free surface can be naturally identified from a subset of particles without requiring additional interface-tracking procedures. As a result, free surfaces are represented sharply and clearly, making SPH particularly suitable for problems involving large interface deformations. However, because particle connectivity changes at every time step, the search for neighboring particles is computationally expensive \cite{npl}. In addition, SPH methods suffer from consistency issues \cite{overview}. Therefore, combining mesh-based and particle-based methods is an appealing strategy for the simulation of free-surface flows.

Hybrid mesh-based and particle-based methods have emerged over the last decade to exploit the complementary strengths of both frameworks. Morrone \cite{fvsph1,fvsph2} employed the finite volume method (FVM) to resolve the bulk flow and SPH to capture free-surface deformation and breaking. Information was exchanged through a designed overlapping region with predefined particle region. Napoli \cite{cfvsph} also coupled FVM and SPH for free-surface flow simulation and showed that the coupled FV-SPH method incurred lower computational cost than the full SPH approach. However, the use of only a single transition layer between the FV and particle domains may not ensure a sufficiently smooth solution transition across the mesh-particle interface and also imposes restrictions on the ratio of mesh size to particle spacing. Zhang \cite{spem} proposed a hybrid method combining SPH and the smoothed finite element method (SFEM) for fluid-structure interaction (FSI) problems. The SFEM formulation for incompressible flow was developed in a Lagrangian framework, and the conversion from elements to particles was irreversible. Consequently, as the simulation progresses, all elements are eventually converted into particles, reducing the advantages of mesh-based methods in terms of computational efficiency and accuracy. Gao \cite{spmh} proposed a smoothed particle-mesh hydrodynamics (SPMH) method for FSI problems, in which mesh-based FVM was applied only in a user-defined mesh region near the structures to improve local accuracy. Since the rest of the domain remained particle-dominated, the overall computational efficiency and accuracy of SPMH were still constrained by the particle region.

In recent years, gas-kinetic schemes (GKS) have made significant progress to the high-fidelity simulation of compressible flows. GKS is based on the integral solution of kinetic equations \cite{xu2001}, which provides an intrinsic advantage: by combining inviscid and viscous fluxes simultaneously from the gas distribution function constructed at cell interfaces. The applicability of GKS extends well beyond the Euler and Navier--Stokes equations to the shallow water equations \cite{zhao-swe, liu-swe}, magnetohydrodynamics \cite{pu-plasma}, and turbulent flow simulations \cite{cao}. Moreover, high-order compact schemes can be naturally developed within the GKS framework by exploiting the time-accurate gas distribution function at cell interfaces \cite{zhao2023direct,ji-hweno,geno, zhao2019compact}.

This study proposes a novel adaptive finite volume-particle method (AFVPM) that synergistically integrates unstructured-mesh FVM and weakly compressible SPH. The proposed framework features three key innovations: (i) an adaptive two-way conversion strategy that dynamically employs FVM in bulk-flow regions to enhance computational efficiency and accuracy, while using SPH in interfacial regions for free-surface capturing; (ii) a robust mesh-particle interface algorithm that enables stable and seamless data communication between mesh and particle regions; and (iii) an isothermal GKS that accommodates gravitational effects and arbitrary equations of state. Numerical results for several benchmark cases demonstrate the accuracy and efficiency of AFVPM.

This paper is structured as follows: Section 2 presents the methodology of SPH formulation and GKS on unstructured meshes. Section 3 introduces the details of the coupled approach and treatment of mesh-particle interfaces in AFVPM. Section 4 applies AFVPM to several benchmark cases and the final section provides the concluding remarks.

\section{Methodology}
In this section, the weakly compressible SPH formulation and the isothermal gas-kinetic scheme for low-speed flows based on FVM framwork are introduced.

\subsection{Weakly compressible smooth particle hydrodynamics} \label{wcsph}
To develop SPH formulation for low-speed flow, the Navier-Stokes equations in Lagrangian form are deduced,
\begin{equation} \label{eq:nslag}
    \left \{
\begin{aligned}
    \frac{D \rho}{D t} &= -\rho \nabla \cdot \mathbf{U}, \\
    \frac{D \mathbf{U}}{D t} &= \frac{\nabla p}{\rho} + \frac{1}{\rho} \nabla \cdot \mathbf{\tau} + \mathbf{G}, \\
    \frac{D \mathbf{r}}{D t} &= \mathbf{U}.
\end{aligned}
    \right.
\end{equation}
where, $\rho,\mathbf{U},p, \mathbf{G}, \mathbf{\tau}$ denote density, velocity, pressure, external force and viscous stress tensor, respectively. And a weakly compressible equation of state is used to relate pressure with density,
\begin{equation} \label{eq:eos-water}
    p=C_{0}^2(\rho - \rho_{0}),
\end{equation}
where, $\rho_0$ is the reference density. To ensure that the variation of density remains within $1\%$, $C_0$ is chosen to satisfy
$$
C_0 \ge 10(|\mathbf{U}|_{max}, \sqrt{p_{max} / \rho}).
$$
where, $|\mathbf{U}|_{max}, p_{max}$ denote the maximum velocity magnitude and pressure in the computational domain, respectively.

The particle discretizations of spatial differentials in Eq. (\ref{eq:nslag}) are
$$
<\nabla \cdot \mathbf{U}>_i = \sum_j (\mathbf{U}_j - \mathbf{U}_i) \cdot \nabla_i W_{ij} V_j,
$$
$$
<\nabla p>_i = \sum_j (p_i + p_j) \nabla_i W_{ij} V_j,
$$
$$
<\nabla \cdot \mathbf{\tau}>_i = \mu \sum_j \pi_{ij} \nabla_i W_{ij} V_j,
$$
$$
\pi_{ij} = 8\frac{(\mathbf{U}_j - \mathbf{U}_i)\cdot \mathbf{r}_{ji}}{||\mathbf{r}_{ji}||^2}.
$$
where, the notation $<>$ denotes the particle approximations in SPH,  $V$ denotes the volume of particles, and $W$ is the smoothing kernel function. C-2 Wendland kernel is used in this paper, defined as
\begin{equation}
    W(q) = \left \{
\begin{aligned}
    &\frac{7}{4\pi h^2}(1-\frac{1}{2}q)^4(2q + 1), &q \le 2,\\
    &0, &\text{elsewhere},
\end{aligned}
    \right.
\end{equation}
where, $q = \frac{||\mathbf{r}_j - \mathbf{r}_i||}{h}$, $h$ is the smoothing length, $h = 1.5\Delta x$ in this paper, $\Delta x$ indicates the initial particle space.

To improve the stability and alleviate the pressure fluctuation of SPH, we introduce the numerical diffusive term $D^{\rho}$ as Gao \cite{gaoapr} to the right-hand side of mass equation,
$$
D^{\rho}_i = 2 \delta h C_0 \sum_j \Phi_{ij} \cdot \nabla_i W_{ij} V_j,
$$
where,
$$
\Phi_{ij} = 2(\rho_j - \rho_i)\frac{\mathbf{r}_j - \mathbf{r}_i}{||\mathbf{r}_j - \mathbf{r}_i||^2},
$$
and $\delta = 0.1$ in this paper.

To regularize the particle distribution, the Particle Shifting Technique (PST) proposed in \cite{alesph} is employed. The velocity deviation arising from irregular particle configurations is expressed as:
\begin{equation}
    \delta \mathbf{U}_i = \text{min}\{ ||\delta \mathbf{U}^*_i||, \frac{U_{max}}{2}\} \frac{\delta \mathbf{U}^*_i}{||\delta \mathbf{U}^*_i||}.
\end{equation}
where, $U_{max}$ denotes the maximum velocity within the computational domain, and the term $\delta \mathbf{U}^*_i$ is evaluated as follows:
$$
\delta \mathbf{U}^*_i = -2h_iU_{max}\sum_j[1+R(\frac{W_{ij}}{W(\Delta x)})^n]\nabla_iW_{ij} V_j,
$$
Following the recommendations in \cite{alesph}, the constants 
$R$ and $n$ are set to 0.2 and 4, respectively.

Above all, the final formulations of the SPH are
\begin{equation}
    \left \{
\begin{aligned}
    \frac{D \rho_i}{D t} &= -\rho_i \sum_j (\mathbf{U}_j - \mathbf{U}_i) \cdot \nabla_i W_{ij} V_j + D^{\rho}_i, \\
    \frac{D \mathbf{U}_i}{D t} &= \frac{1}{\rho_i}\sum_j (p_i + p_j) \nabla_i W_{ij} V_j + \frac{1}{\rho_i} \mu \sum_j \pi_{ij} \nabla_i W_{ij} V_j + \mathbf{G}, \\
    \frac{D \mathbf{r}_i}{D t} &= \mathbf{U}_i + \delta \mathbf{U}_i.
\end{aligned}
    \right.
\end{equation}

The fast free-surface detection technique proposed by Marrone \cite{surfDetect} is adapoted to detect the particles on free surface.

\subsection{Finite volume gas-kinetic scheme for isothermal flow} \label{fvgks}
Gas-kinetic solver is constructed from the BGK equation,
\begin{equation}\label{bgk}
	f_t+\mathbf{u}\cdot \nabla_{\mathbf{x}} f + \mathbf{G} \cdot \nabla_{\mathbf{u}} f = \frac{g-f}{\tau},
\end{equation}
where, $f = f(\mathbf{x}, t, \mathbf{u})$ is the gas distribution function, which is a function of space $\mathbf{x}$, time $t$ and particle velocity $\mathbf{u}$, $\mathbf{G}$ is the external force, here we consider the gravity force. $\tau$ is the relaxation time from $f$ to its equilibrium state $g$. The equilibrium state $g$ considers Maxwellian distribution,
\begin{equation}
	g = \rho (\frac{\lambda}{\pi})e^{-\lambda (\mathbf{u} - \mathbf{U})^2}, \nonumber
\end{equation}
where, $\rho$ and $\mathbf{U}$ are density and velocity respectively, $\lambda = \frac{1}{2T_0}$ where $T_0 = C_0^2$ for isothermal flows.

Due to the conservation of mass and momentum during particle collision process, the collision term in Eq. (\ref{bgk}) should satisfy the compatibility condition,
\begin{equation} \label{comp_cond}
	\int \frac{g-f}{\tau} \psi \mathrm{d} \Xi = 0,
\end{equation}
where $\psi = (1, \mathbf{u})^{T}$, $\mathrm{d} \Xi = \mathrm{d}u\mathrm{d}v$.
The conservative variables and fluxes can been obtained from the gas distribution function,
\begin{equation} \label{consvar}
	\mathbf{W} = \int f \psi \mathrm{d} \Xi,
\end{equation}
\begin{equation} \label{flux}
	\mathbf{F} = \int f \psi \mathbf{u} \mathrm{d} \Xi,
\end{equation}
where, $\mathbf{W}=\{\rho, \rho \mathbf{U}\}^{T}$ is the conservative variables and $\mathbf{F}$ is the corresponding flux.

\subsubsection{Finite volume formulation}
Taking the moments of the BGK model of Eq. (\ref{bgk}) in phase space and integrating over a finite volume $\Omega_i$ yields:
\begin{equation}
	\int_{\Omega_i}\int (f_t+\mathbf{u}\cdot \nabla_{\mathbf{x}} f + \mathbf{G} \cdot \nabla_{\mathbf{u}} f)\psi \mathrm{d}\Xi\mathrm{d} V = \int_{\Omega_i}\int \frac{g-f}{\tau}\psi \mathrm{d}\Xi\mathrm{d} V. \nonumber
\end{equation}
By applying the relations given in Eqs. (\ref{comp_cond}-\ref{flux}), the integral form is obtained as follows:
\begin{equation} \label{int_form}
	\int_{\Omega_i} \mathbf{W}_t \mathrm{d} V + \int_{\Omega_i} \nabla \cdot \mathbf{F} \mathrm{d} V + \int_{\Omega_i} \mathbf{s} \mathrm{d} V= 0. \nonumber
\end{equation}
The above integral equation is then discretized using the FVM, resulting in:
\begin{equation} \label{semi_discrete}
	\frac{\mathrm{d} \overline{\mathbf{W}}_i}{\mathrm{d} t} = -\frac{1}{|\Omega_i|}\int_{\Omega_i} \nabla \cdot \mathbf{F} \mathrm{d} V - \frac{1}{|\Omega_i|} \int_{\Omega_i} \mathbf{s} \mathrm{d} V.
\end{equation}
and applying Gauss's theorem, the semi-discrete form in Eq. (\ref{semi_discrete}) can be rewritten as:
\begin{equation} \label{res}
	\frac{\mathrm{d} \overline{\mathbf{W}}_i}{\mathrm{d} t}
	= \mathcal{L}(\mathbf{W}_i)
	= -\frac{1}{|\Omega_i|} \sum_{j=1}^{N_f}\int_{\Gamma_{ij}} \mathbf{F} \cdot \mathbf{n}_j \mathrm{d}s - \frac{1}{|\Omega_i|} \int_{\Omega_i} \mathbf{s} \mathrm{d} V,
\end{equation}
where, $|\Omega_i|$ is the volume of the mesh cell, $\partial \Omega_i$ is the boundary of the control volume, which is expressed as:
\begin{equation}
	\partial \Omega_i = \bigcup_{j=1}^{N_f}\Gamma_{ij}. \nonumber
\end{equation}
Here $\Gamma_{ij}$ indicates the $j-th$ interface of cell $\Omega_i$ and $N_f$ is the number of cell interfaces.
The surface integral of the fluxes and volume integral of source $\mathbf{S}$ are approximated by:
\begin{equation} \label{eq:faceInt}
	\int_{\Gamma_{ij}} \mathbf{F} \cdot \mathbf{n}_j \mathrm{d}s = |\Gamma_{ij}|\mathbf{F}(\mathbf{x}_{j},t)\cdot \mathbf{n}_{j},
\end{equation}
$$
\frac{1}{|\Omega_i|} \int_{\Omega_i} \mathbf{s} \mathrm{d} V = \overline{\mathbf{s}}_i.
$$
where, $|\Gamma_{ij}|$ is the area of the cell interface, $\mathbf{x}_{j}$ is the center and $\mathbf{n}_{j}$ is the unit normal vector.
\subsubsection{Isothermal gas-kinetic solver} \label{gks_2nd}
To evaluate the fluxes across the mesh interface in Eq. (\ref{eq:faceInt}), an isothermal gas-kinetic scheme is developed based on the integral solution of Eq. (\ref{bgk}) derived by Xu \cite{xu2001},
\begin{equation} \label{integral_sol}
	f(\mathbf{x}, t, \mathbf{u}) = \frac{1}{\tau}\int_{0}^{t} g(\mathbf{x}', t', \mathbf{u}')e^{-\frac{t-t'}{\tau_n}}\mathrm{d}t'+e^{\frac{-t}{\tau_n}}f_0(\mathbf{x}_0, 0, \mathbf{u}_0),
\end{equation}
where, $\mathbf{x} = \mathbf{x}' + \mathbf{u} (t-t') + \frac{1}{2}\mathbf{G} (t-t')^2, \mathbf{u} = \mathbf{u}' + \mathbf{G}  (t-t')$ are the trajectory and velocity of the particle motion, let $\mathbf{x} = \mathbf{0}$ is the location of cell interface,
$f_0$ is the gas distribution function at the beginning of each time step ($t=0$), $\mathbf{x}_0$ is the initial position of the concerned particle.
For the gas-kinetic solver, the equilibrium term $g(\mathbf{x}', t', \mathbf{u}')$ and initial distribution function $f_0(\mathbf{x}_0, \mathbf{u})$ need to be constructed.
To construct a time-accurate gas distribution function at the cell interface, the following notations are introduced firstly,
\begin{equation}
	\mathbf{a} = \frac{1}{g}\frac{\partial g}{\partial \mathbf{x}}, A = \frac{1}{g}\frac{\partial g}{\partial t}, \nonumber
\end{equation}
where, $\mathbf{a}$ and $A$ are the spatial and temporal derivative of the equilibrium distribution function. Apply differentials to Maxwellian distribution function,
\begin{equation} \label{eq:g_x}
    \mathbf{a} = a_1 + a_2 u + a_3 v,
\end{equation}
where,
\begin{equation} \label{eq:exp_a}
    \begin{aligned}
        a_3 &= 2 \lambda \partial V, \\
        a_2 &= 2 \lambda \partial U, \\
        a_1 &= \frac{\partial \rho}{\rho} - a_2 U - a_3 V.
    \end{aligned}
\end{equation}
$A$ can be determined in a similar way as Eq. (\ref{eq:g_x}),
$$
A = A_1 + A_2 u + A_3 v,
$$
which satisfies
$$
\int A g \psi \mathrm{d}\Xi = -\int \mathbf{a} \cdot \mathbf{u}g \psi \mathrm{d} \Xi = \partial_t \mathbf{W}.
$$
from the compatibility condition Eq. (\ref{comp_cond}).

According to Taylor expansion, the equilibrium part inside the integration in Eq. (\ref{integral_sol}) can be expressed as,
\begin{equation} \label{g}
	\begin{aligned}
		g(\mathbf{x}', t', \mathbf{u}') = g(\mathbf{0}, 0, \mathbf{u})(1-\mathbf{a} \cdot \mathbf{u}(t-t') +2\lambda\mathbf{G} \cdot (\mathbf{u} - \mathbf{U})(t - t')+ At').
	\end{aligned}
\end{equation}
The last part in Eq. (\ref{integral_sol}) is recast by
\begin{equation}\label{f0}
	\begin{aligned}
		f_0(\mathbf{x}_0, 0, \mathbf{u}_0)
		&=g^{l}(\mathbf{0}, 0, \mathbf{u})(1-(t+\tau)\mathbf{a}^{l/r} \cdot \mathbf{u} +2 \lambda \mathbf{G} \cdot (\mathbf{u} - \mathbf{U})(t + \tau)- \tau A^{l/r})H(u) \\
		&+g^{r}(\mathbf{0}, 0, \mathbf{u})(1-(t+\tau)\mathbf{a}^{l/r} \cdot \mathbf{u} +2\lambda \mathbf{G} \cdot (\mathbf{u} - \mathbf{U})(t + \tau)- \tau A^{l/r})(1-H(u)).
	\end{aligned}
\end{equation}
Substituting Eq. (\ref{g}) and Eq. (\ref{f0}) into the integral solution Eq. (\ref{integral_sol}), the time-accurate distribution function at the interface can be obtained,
\begin{equation} \label{f_inter}
	\begin{aligned}
		f(\mathbf{0}, t, \mathbf{u}) &= C_1g^c+C_2\mathbf{a}^c\cdot \mathbf{u} g^c -C_2 2  \lambda^c \mathbf{G}\cdot (\mathbf{u} - \mathbf{U}^c)+ C_3A^cg^c \\
		&+ C_4[g^lH(u) + g^r(1-H(u))] \\
		&+ C_5[g^l\mathbf{a}^l\cdot \mathbf{u}H(u) + g^r\mathbf{a}^r\cdot \mathbf{u}(1-H(u))] \\
		&- C_5[g^l 2 \lambda^l \mathbf{G} \cdot (\mathbf{u} - \mathbf{U}^l) H(u) + g^r 2 \lambda^r \mathbf{G} \cdot (\mathbf{u} - \mathbf{U}^r)(1-H(u))] \\
		&+ C_6[g^lA^lH(u) + g^rA^r(1-H(u))],
	\end{aligned}
\end{equation}
with the coefficients $C_1 = 1-e^{-t/\tau_n}$, $C_2 = (t+\tau)e^{-t/\tau_n}-\tau$, $C_3 = t-\tau+\tau e^{-t/\tau_n}$, $C_4 = e^{-t/\tau_n}$, $C_5 = -e^{-t/\tau_n}(\tau + t)$, $C_6 = -\tau e^{-t/\tau_n}$.

Once the gas distribution function at the mesh interface is determined, the fluxes can be evaluated by Eq. (\ref{flux}). 
It is noteworthy that the standard GKS assumes the ideal gas equation of state (EOS), i.e., $p = \rho T_0$. To apply the GKS to fluids governed by other types of EOS, such as $p = p(\rho)$ in Eq. (\ref{eq:eos-water}), an additional pressure term must be incorporated into the numerical flux, as follows:
\begin{equation}
    \mathbf{F}_{j+1/2} = \int f(\mathbf{0}, t, \mathbf{u}, \xi) \psi \mathbf{u} \mathrm{d} \Xi +
    \left \{
\begin{aligned}
    &0 \\
    p(\rho^c) &- \rho^c T_0 \\
    &0
\end{aligned}
    \right \}.
\end{equation}

For viscous flow, the physical collision time is determined by
\begin{equation}
	\tau = \frac{\mu}{p},  \nonumber
\end{equation}
where, $\mu$ is the dynamic viscosity. In this paper, we add some numerical dissipation by modifying the numerical collision time to eliminate the numerical oscillations in the following way,
\begin{equation}
	\tau_n = \frac{\mu}{p} + C\frac{|p_l - p_r|}{|p_l + p_r|}\Delta t,  \nonumber
\end{equation}
where, $\Delta t$ is the time step, $p_l$ and $p_r$ are the pressure on the left and right side of the interface, $C = 0.01$ in this paper.

In addition to the conservative variables on the left and right sides of the cell interfaces, their spatial gradients are also required to evaluate the $\mathbf{a}$ and $A$ terms in Eq. (\ref{f_inter}) using Eq. (\ref{eq:exp_a}). To achieve second-order accuracy, the widely used Green-Gauss method is employed to compute these gradients.

\subsection{Temporal discretization}
For simplification, the forward Euler method is used to update the mesh and particle fields in the following way.

Mesh cells:
$$
\overline{\mathbf{W}}^{n+1}_j=\overline{\mathbf{W}}^{n}_j + \Delta t \mathcal{L}(\overline{\mathbf{W}}^{n}_j),
$$

SPH particles:
\begin{equation}
    \left \{
    \begin{aligned}
        \rho^{n+1}_i &= \rho^{n}_i + \Delta t \frac{D\rho_i}{Dt}, \\
        \mathbf{U}^{n+1}_i &= \mathbf{U}^{n}_i + \Delta t \frac{D\mathbf{U}_i}{Dt}, \\
        \mathbf{r}^{n+1}_i &= \mathbf{r}^{n}_i + \Delta t \frac{D\mathbf{r}_i}{Dt}.
    \end{aligned}
    \right.
    \nonumber
\end{equation}
where, $\Delta t$ is the computational time step which takes the minimum values determined by particle and mesh domain,
$$
\Delta t = \mathrm{CFL} \cdot \text{min}\{\frac{\Delta x}{C_0 + ||\mathbf{U}_j||}, \frac{\Delta x^2}{4\nu} \},
$$
where, $\mathrm{CFL}$ is the Courant-Friedriches-Lewy number, which takes $\mathrm{CFL} = 0.3$ in this work, $\Delta x$ is the initial  particle spacing which is identical to the background mesh size, $\nu$ is the kinematic viscosity.

\section{AFVPM: Coupled approach and treatment of mesh-particle interface} \label{afvpm}
\begin{figure}[!htbp]
	\centering
	\includegraphics[width = 0.6\columnwidth, clip]{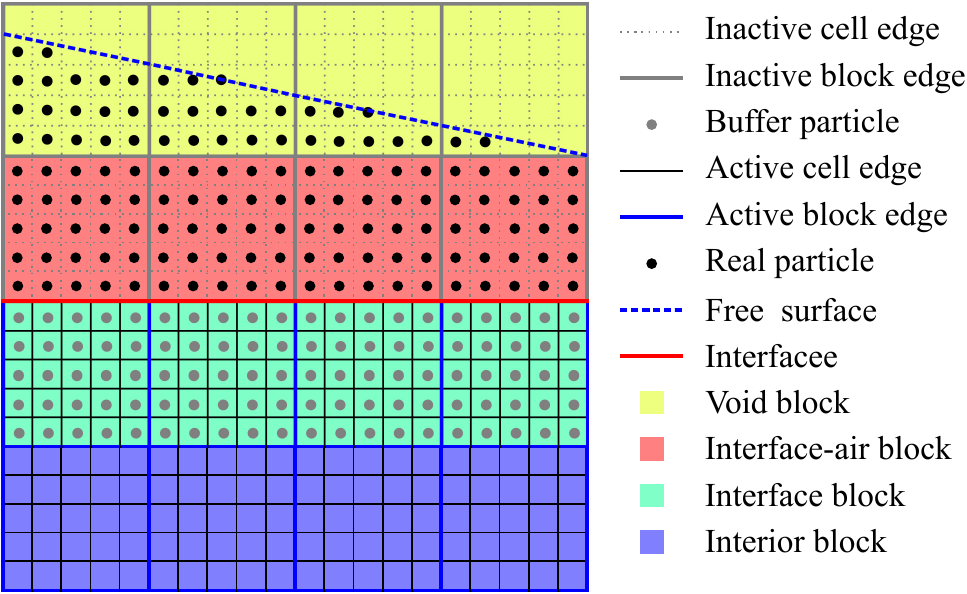}
	\caption{Domain partition of AFVPM}
	\label{fig:FVPM}
\end{figure}

 In AFVPM, the computational domain is composed of Eulerian mesh region and Lagrangian particle region, as shown in Figure \ref{fig:FVPM}. To achieve the conversion from particle region to mesh region, a background mesh is needed, while only the active parts occupied by fluid need to be updated during the iteration. To simplify the conversion between particle region and mesh region and regularize the field, the background mesh is divided into Cartesian mesh blocks, which are the minimum conversion units in AFVPM rather than each mesh cell. The size of mesh block should satisfy,
 $$
 \mathrm{DX} \ge \text{max}\{2kh,2\Delta x \},
 $$
 where, $kh$ is the smoothing length of the kernel function, $\Delta x$ is the initial mesh / particle size. This condition ensures that the free surface particles are never affected by the active mesh cells, and there are at least two layers of mesh cells in each block. In this work, the block size is taken as $\mathrm{DX} = 10 \Delta x$.

 The background mesh blocks are divided into four categories: Interior, interface-mesh, interface-air and void. The interior blocks denote the blocks filled by fluid, and far from the free surface, the interface-mesh blocks are also filled by fluid but neighboring the particle region, both interior blocks and interface-mesh blocks are collectively called activated mesh blocks and computed by GKS introduced in Section \ref{fvgks}, the interface-particle and void blocks are occupied by particles and collectively called deactivated blocks, the flow fields are computed by SPH method. We denote the mesh cells in interface-air blocks as buffer cells, to evaluate the flux across the mesh-particle interfaces, the field variables in buffer cells are needed, which will be introduced in Section \ref{sec:buffercell}.

There are two types of particles in the computational domain: real particles and buffer particles. The particles located in interface-air and void blocks are called real particles and updated by the SPH formula introduced in Section \ref{wcsph}, besides these particles, to complete the support of kernel function of the real particles, a set of buffer particles are generated in the interface-mesh blocks, and updated with the real particles to ensure uniform distribution of particle near mesh-particle interface, and seamless data communication across the interface, as introduced in Section \ref{sec:bufferpar}.

\subsection{Update of buffer particles} \label{sec:bufferpar}
To complete the support of kernel function of real particles, a set of buffer particles are generated initially at the center of the mesh cells in interface-mesh blocks. The initial field variables $\phi$, i.e. density and velocity, of the buffer particles are set from the cell-averaged conservative variables,
\begin{equation} \label{eq:buffer1}
    \begin{aligned}
        A_p = \Omega_m, \\
        \rho_p = \overline{\rho}_m, \\
        \mathbf{U}_p = \frac{\overline{\mathbf{\rho U}}_m}{\overline{\rho}_m},
    \end{aligned}
\end{equation}
where, the subscript $p$ indicates the field variables of the particles, $m$ denotes the variables of mesh cell.

To ensure uniform particle distribution near the mesh-particle interface and seamless data communication across the interface, the buffer particles are also updated. Their variables are interpolated from the evolved states of neighboring real particles and active mesh cells,
\begin{equation} \label{eq:bufferPar}
\begin{aligned}
    \phi^{n+1}_i &= \frac{\sum_j^{Np} \phi^{n+1}_j W_{ij} V_j + \sum_j^{Nc}\overline{\phi}^{n+1}_j W_{ij} \Omega_j}{\sum_j^{Np} W_{ij} V_j + \sum_j^{Nc} W_{ij} \Omega_j}, \\
    \mathbf{r}^{n+1}_i &= \mathbf{r}^{n}_i + \frac{\mathbf{U}^{n}_i + \mathbf{U}^{n + 1}_i}{2} \Delta t.
\end{aligned}
\end{equation}
where, $Np$ is the number of neighboring real particles, $Nc$ is the number of neighboring active mesh cells.

\begin{figure}[!htbp]
	\centering
	\includegraphics[width = 0.6\columnwidth, clip]{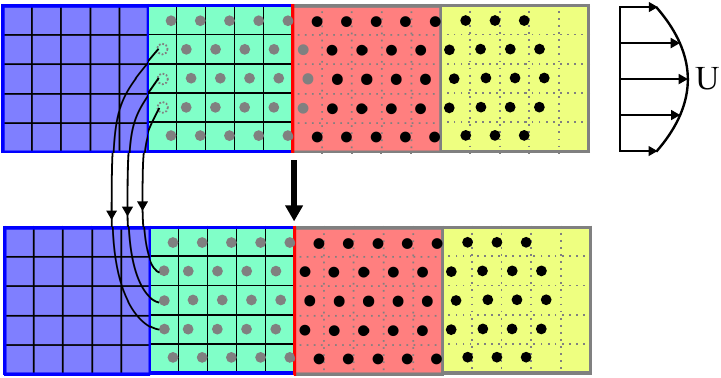}
	\caption{Conversion of buffer and real particles: buffer particles are converted to real particles when entering particle regions and new buffer particles are generated}
	\label{fig:buffer}
\end{figure}

As demonstrated in Figure \ref{fig:buffer}, when the ghost particles enter real particle region, they will be converted to real particles and evolve by SPH formula, if real particles enter interface-mesh blocks, the real particles will be converted to buffer particles. If there is no neighboring buffer or real particles within the range of $\Delta x$ from a mesh cell in interface-mesh blocks, one buffer particle will be supplemented at the cell center by Eq. (\ref{eq:buffer1}).

\subsection{Update of buffer cells} \label{sec:buffercell}
Field variables $\mathbf{W}$ and their gradients in buffer cells are needed to evaluate the flux across the mesh-particle interface as introduced in Section \ref{gks_2nd}. Firstly, the field variables $\phi$ at the center of buffer cells are estimated as the cell-averaged values by,
\begin{equation}\label{eq:bufferCell}
    \phi_i=\frac{\sum^{N_p}_j \phi_j W_{ij} V_j + \sum^{N_b}_j \phi_j W_{ij} V_j}{\sum^{N_p}_j W_{ij} V_j + \sum^{N_b}_j W_{ij} V_j},
\end{equation}
where, $N_p$ is the number of neighboring real particles, $N_b$ is the number of neighboring buffer particles. Then Green-Gauss method is used to construct the gradients of field variables as the interior cells.

\begin{figure}[!htbp]
	\centering
	\includegraphics[width = 0.6\columnwidth, clip]{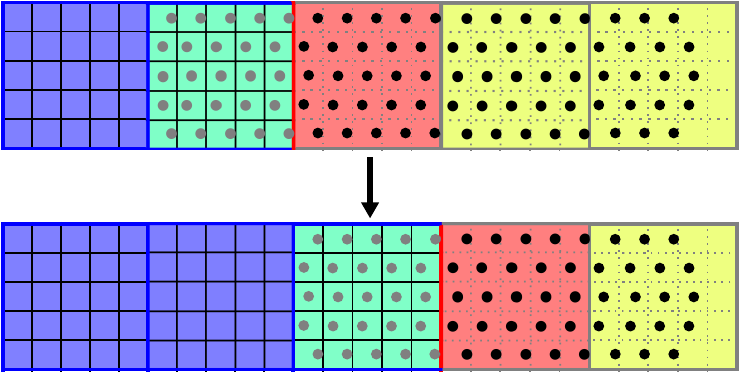}
	\caption{Conversion from interface-air blocks to interface-mesh blocks: interface-air blocks are converted to interface-mesh blocks when far from the free surface}
	\label{fig:partomesh}
\end{figure}

If the distance between an interface-air block and the free surface is greater than the size of Cartesian mesh block $\mathrm{DX}$, the interface-air block will be converted to an interface-mesh block and the real particles in it will be converted to buffer particles, as shown in Figure \ref{fig:partomesh}.

\subsection{Two-way conversion strategy between mesh and particle region}
To achieve dynamics conversion between Eulerian mesh region and Lagrangian particle region, a two-way conversion strategy involving four mechanisms is proposed:
\begin{enumerate}
	\item Mechanism \Rmnum{1}: If buffer particles enter interior mesh blocks, the particles will be deleted directly.
	\item Mechanism \Rmnum{2}: If a deactivated mesh block is more than $\mathrm{DX}$ distance away from the free surface, this mesh block will be activated and the field variables of the mesh cells inside will be determined by Eq. (\ref{eq:bufferCell}) from the neighboring real and buffer particles, and the real particles inside the block will be deleted.
	\item Mechanism \Rmnum{3}: If an interface-mesh block is within $\mathrm{DX}$ distance from the free surface, the mesh block will be deactivated and the buffer particles inside the block will be converted to real particles.
	\item Mechanism \Rmnum{4}: If the buffer particles enter the real particle region, the buffer particles will be activated to real particles and evolved by SPH formula. If the real partilces enter the mesh region, these will be converted to buffer particles and updated by Eq. (\ref{eq:bufferPar}).
\end{enumerate}
Above all, the main procedures of AFVPM are demonstrated as follows.
\begin{algorithm}
    \caption{Algorithm of AFVPM}
    \label{alg:afvpm}
    \begin{algorithmic}
        \STATE Initialize: Mesh reading and partition
        \FOR{$k=1,\cdots,N_{max}$}
            \STATE Calculate time step
            \STATE Mechanism \Rmnum{3}
            \STATE Update mesh block type
            \STATE Mechanism \Rmnum{2}
            \STATE Update mesh block type
            \STATE Generate buffer particle in new interface-mesh blocks by Eq. (\ref{eq:buffer1})
            \STATE Mechanism \Rmnum{4}
            \STATE Mechanism \Rmnum{1}
            \STATE Neighboring particle searching for real particles
            \STATE Set boundary conditions for real particles
            \STATE Calculate the $\frac{D\rho_i}{Dt},\frac{D\mathbf{U}_i}{Dt}$ for real particles
            \STATE Set boundary conditions for activated mesh cells
            \STATE Calculate fluxes for activated mesh cells
            \STATE Update to next time level
        \ENDFOR
    \end{algorithmic}
\end{algorithm}

\section{Numerical examples} \label{sec:NumExp}
The following examples demonstrate the performance of AFVPM. A hydrostatic still-water tank is first presented, showing good agreement with the analytical solution. The conventional dam breaking problem is then simulated, with results agreeing well with both experimental data and reference solutions. More complex cases, including ship cruising, body entry, and water filling problems, are also examined to further verify the performance of AFVPM, where consistently smooth and accurate results are obtained.
\subsection{Still tank problem}
The still tank case is widely used to test the accuracy and stability of algorithm for free surface flow. The computational domain is $(x,y) \in [0,1]\times [0,1]$ with the water height of $H=0.5$.

\begin{figure}[!htbp]
	\centering
	\subfigure{\includegraphics[width = 0.62\columnwidth, trim = 10 5 10 5, clip]{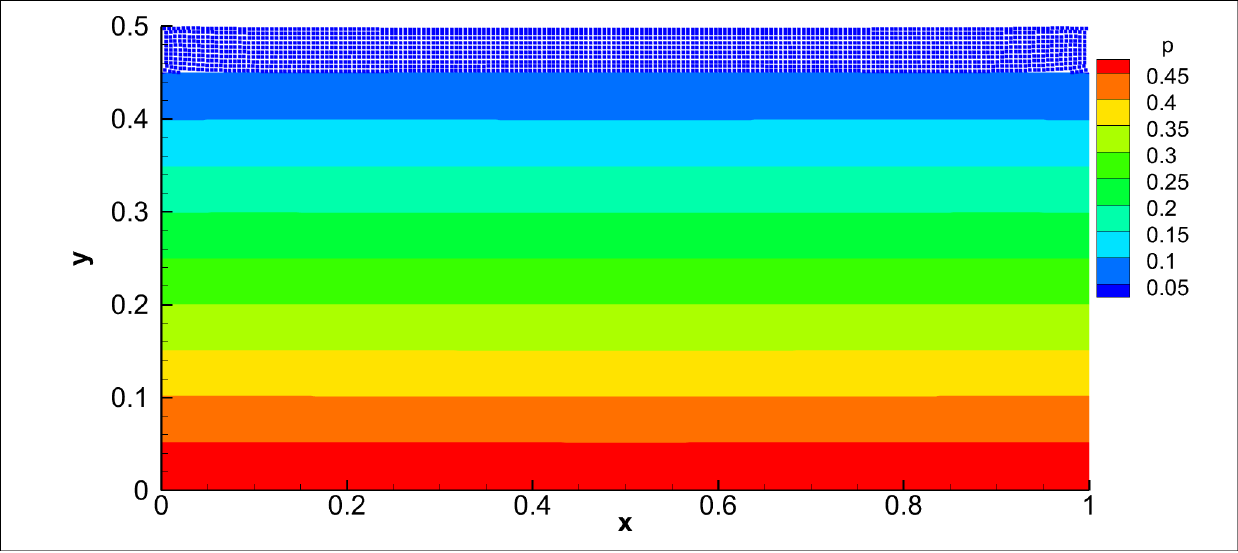}} \hspace{5pt}
	\subfigure{\includegraphics[width = 0.34\columnwidth, trim = 10 10 10 10, clip]{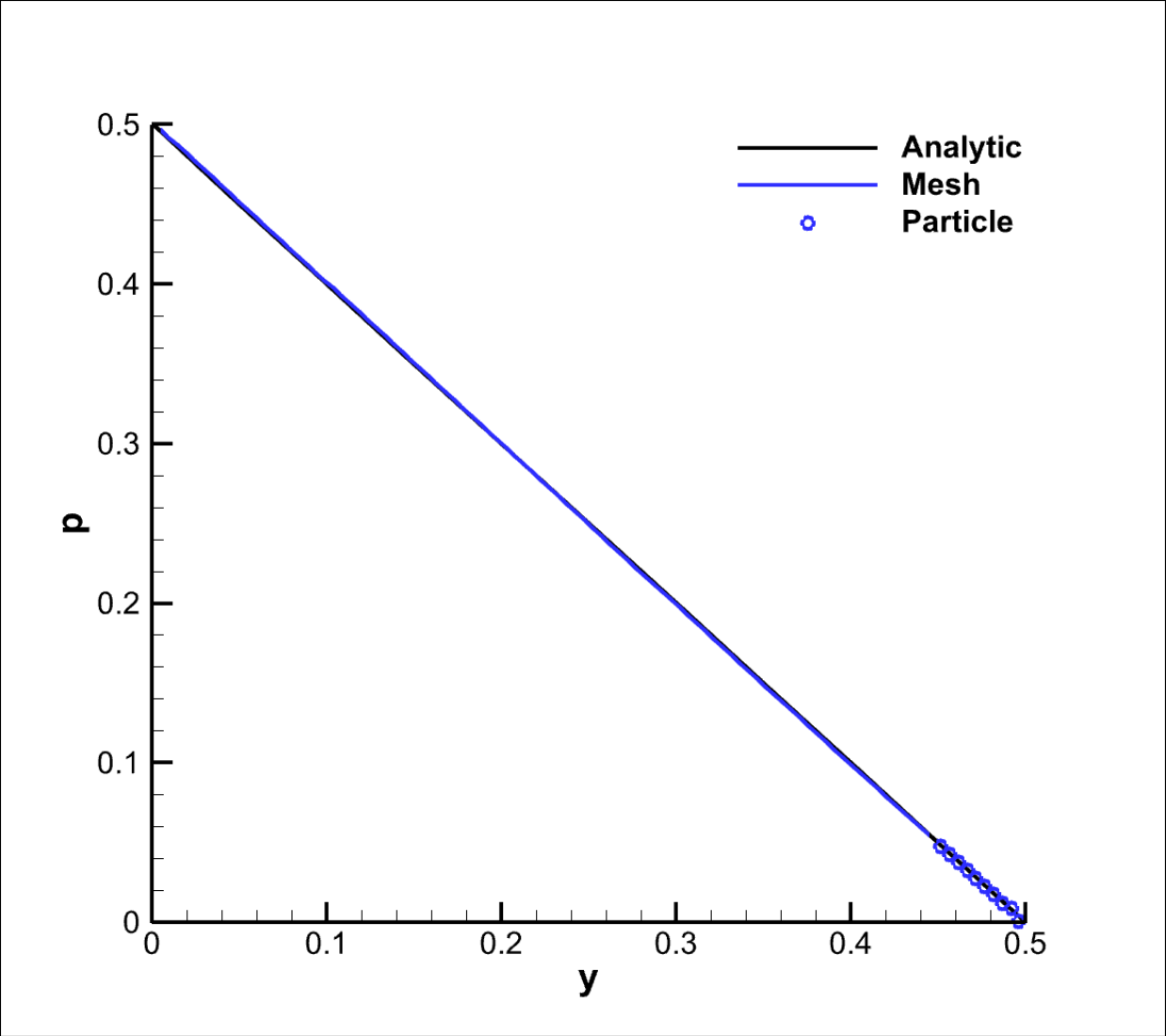}}
	\caption{Pressure distribution of still tank problem}
	\label{fig:stillTank}
\end{figure}

The initial particle space and background mesh size are $\Delta x = 1 / 200$, and the boundaries are set as slip walls, the particles on free surface are detected and set the pressure as $p = 0$ at each iteration. The reference density $\rho_0 = 1$, the artificial sound speed takes $C_0 = 15.0$ and the gravity force is $G = -1$ in y-direction. The pressure contour and distribution along the centerline are shown in Figure \ref{fig:stillTank}, where the numerical result is consistent with the analytical solution.

\subsection{Dam breaking problem}
Now we consider the collapse of a liquid column to verify the dynamics conversion of mesh and particle region. The computational domain is $(x,y) \in [0,5.366]\times [0,3]$, and the water phase covers a rectangle region of $L = 2$ and $H = 1$.

The initial particle space / mesh size is $\Delta x = 1 / 100$, thus a total of 20000 particles / mesh cells are generated initially. The boundaries are all set as slip walls and the particles on free surface are detected and set with $p=0$ at each iteration. The reference density is $\rho_0 = 1$, the gravity force is $G = -1$ in y-direction and sound speed is set as $C_0 = 15\sqrt{2|G|H}$, the dynamics viscosity of $\mu = 3.54\times10^{-3}$ gives a Reynolds number $Re = 400$.

\begin{figure}[!htbp]
	\centering
    \subfigure{\includegraphics[width = 0.8\columnwidth, clip]{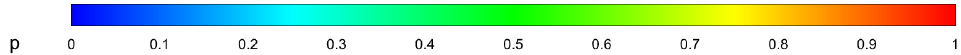}} \\
	\subfigure{\includegraphics[width = 0.48\columnwidth, trim = 30 5 60 10, clip]{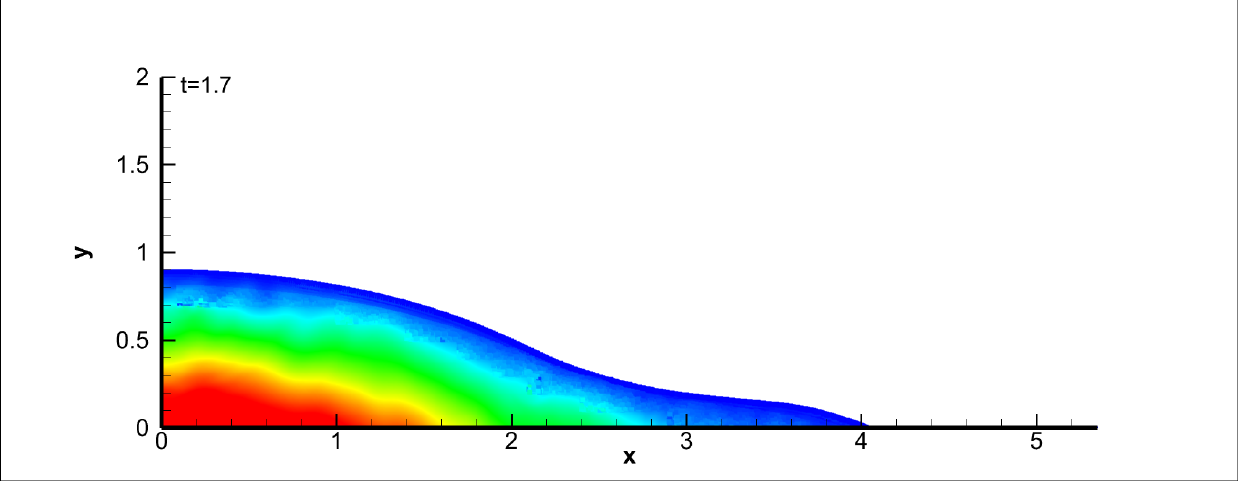}} \hspace{5pt}
	\subfigure{\includegraphics[width = 0.48\columnwidth, trim = 30 5 60 10, clip]{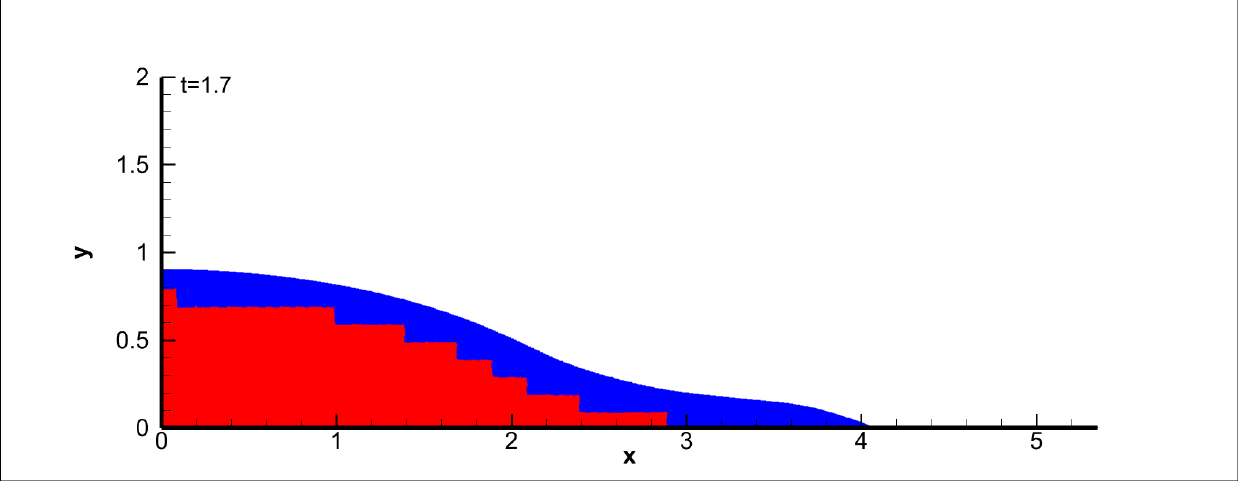}} \\
	\subfigure{\includegraphics[width = 0.48\columnwidth, trim = 30 5 60 10, clip]{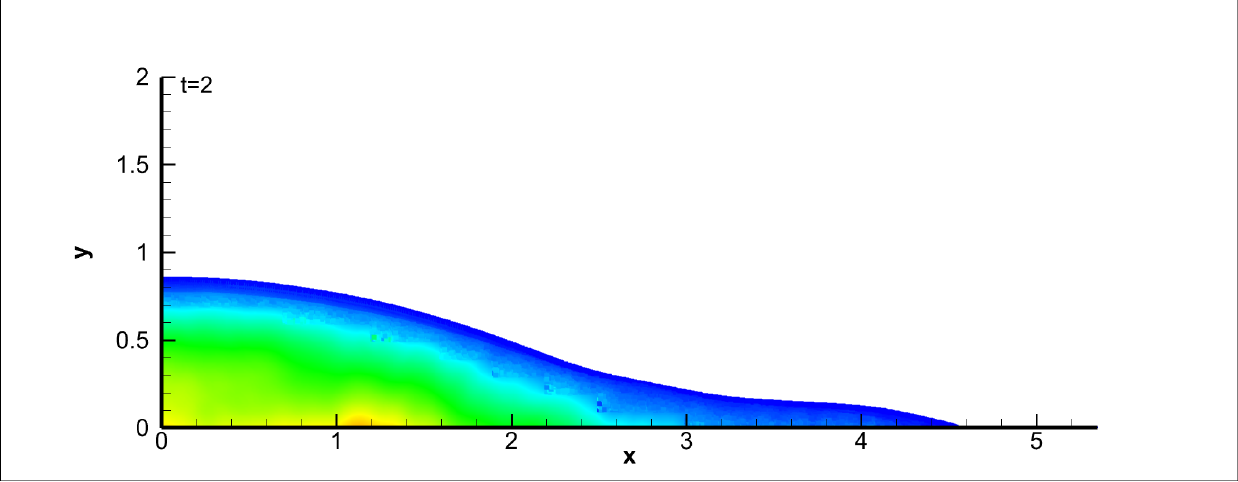}} \hspace{5pt}
	\subfigure{\includegraphics[width = 0.48\columnwidth, trim = 30 5 60 10, clip]{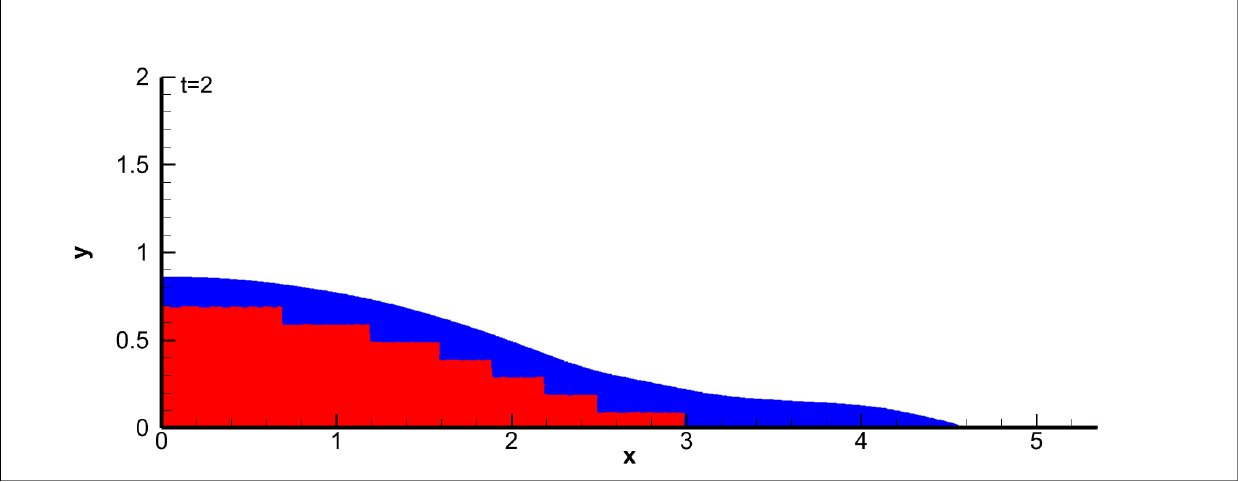}} \\
	\subfigure{\includegraphics[width = 0.48\columnwidth, trim = 30 5 60 10, clip]{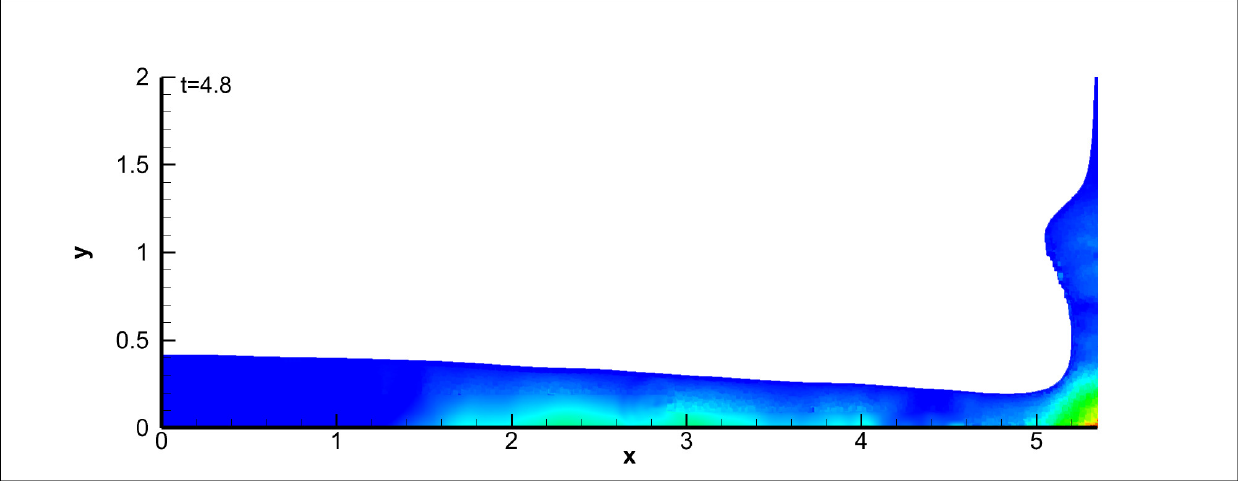}} \hspace{5pt}
	\subfigure{\includegraphics[width = 0.48\columnwidth, trim = 30 5 60 10, clip]{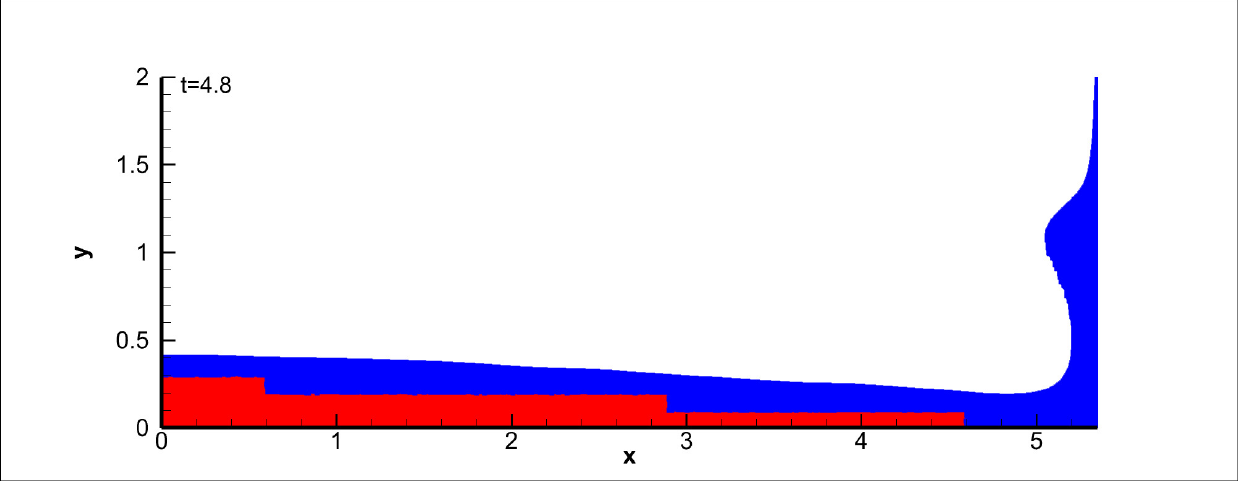}} \\
    \subfigure{\includegraphics[width = 0.48\columnwidth, trim = 30 5 60 10, clip]{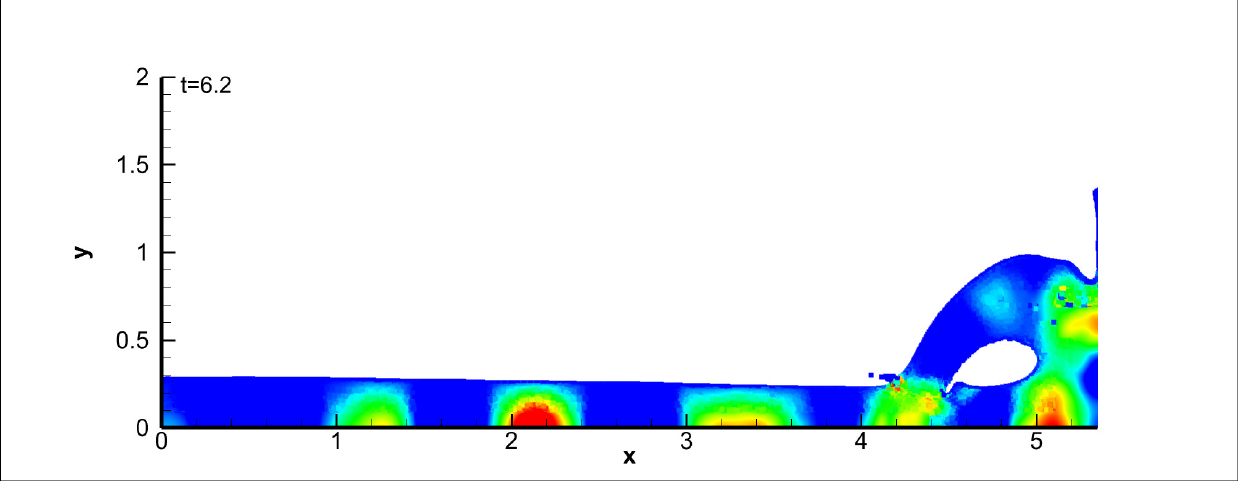}} \hspace{5pt}
	\subfigure{\includegraphics[width = 0.48\columnwidth, trim = 30 5 60 10, clip]{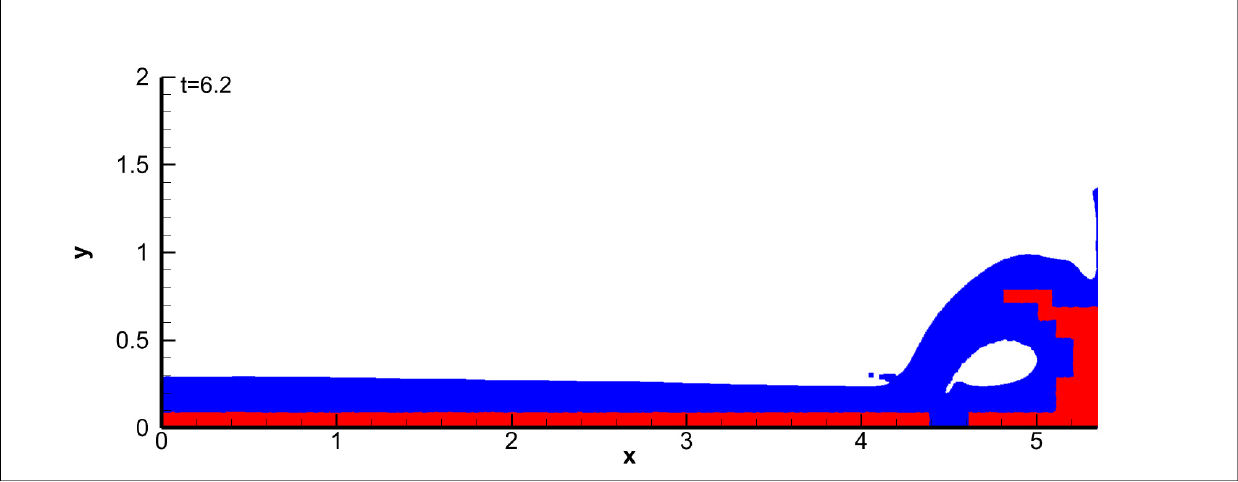}} \\
    \subfigure{\includegraphics[width = 0.48\columnwidth, trim = 30 5 60 10, clip]{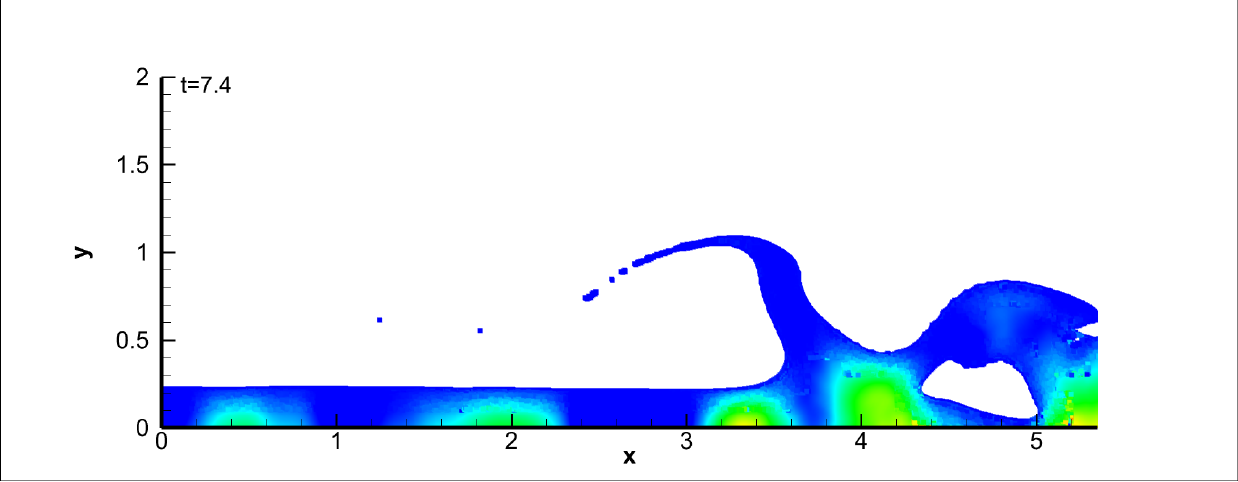}} \hspace{5pt}
	\subfigure{\includegraphics[width = 0.48\columnwidth, trim = 30 5 60 10, clip]{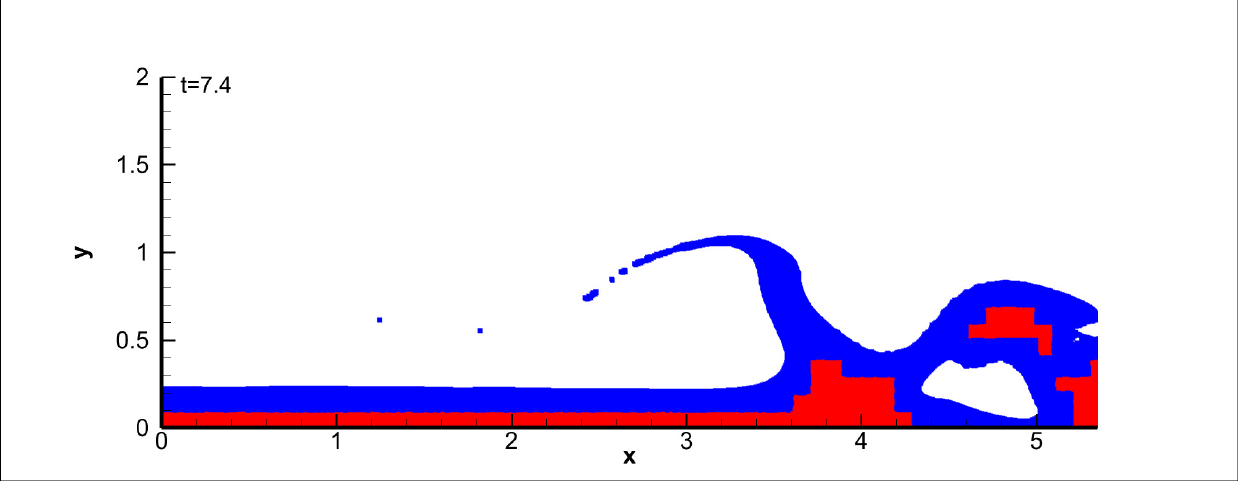}} \\
	\caption{Numerical results of dam breaking problem at $T = 1.7, 2, 4.8, 6.2, 7.4$. Left column: pressure, right column: mesh-particle partitions (red: mesh region, blue: particle region)}
	\label{fig:damBreak}
\end{figure}

\begin{figure}
    \centering
    \includegraphics[width=0.8\linewidth, trim = 3 3 3 3, clip]{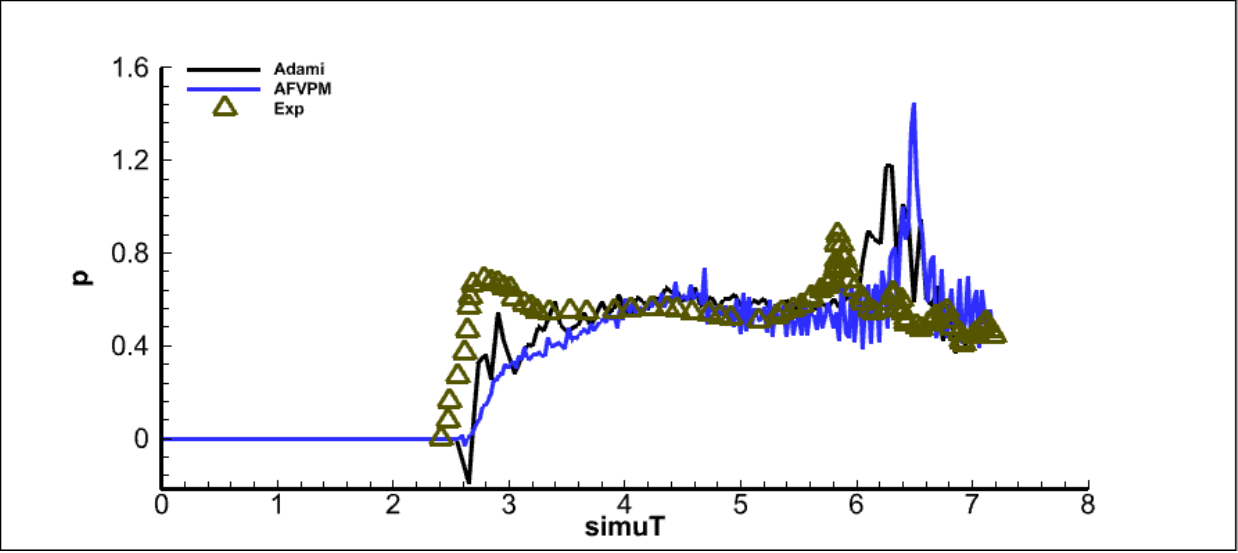}
    \caption{Pressure evoluation at $y=0.2H$ of the right wall}
    \label{fig:damBreak-exp}
\end{figure}

The contour of pressure and partition of mesh and particle region at computational time $T = 1.7, 2, 4.8, 6.2, 7.4$ are shown in Figure \ref{fig:damBreak}. The red regions in Figure \ref{fig:damBreak} are the activated mesh blocks solved by GKS and blue regions are deactivated mesh blocks solved by SPH method. The computational domain can be divided into mesh and particle regions adaptively with the moving free surface. The pressure history at $\frac{y}{H}=0.2$ at the right wall every 500 iterations is shown in Figure \ref{fig:damBreak-exp}, the numerical results by AFVPM show good agreement with the experimental data \cite{damBreak-exp} and results obtained by Adami \cite{adami}.

\subsection{Ship cruising problem}
\begin{figure}
    \centering
    \includegraphics[width=0.7\linewidth, clip]{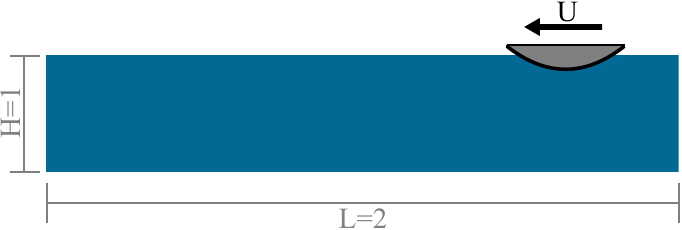}
    \caption{Schematic for ship cruising problem}
    \label{fig:shipGeom}
\end{figure}

A parabolic-shape ship cruising on initially still ocean surface is tested by AFVPM to verify the capacity of dealing with free surface flow with moving body. The computational domain is shown in Figure \ref{fig:shipGeom}, the ship shape is described by a parabola with the formula,
$$
y=\frac{B}{A^2}((x-x_0)^2 - A^2) + y_0,
$$
where, $(x_0, y_0)$ is the initial center of the upper deck of the ship, which takes $(4,1.05)$ in this case, $B = 0.2$ is the maximum depth, $A = 0.5$ is the half length of the deck, which means there is 0.05 height out of the water and 0.15 height below the water surface. The total length of the domain is $L = 5.366$, the initial depth of water is $H = 1$, the initial mesh / particle size is taken as $\Delta x = 1 / 100$, thus a total of 54000 particles / mesh cells are generated initially, and the ship cruises at a speed $U$, defined by
\begin{equation}
    U =
    \left \{
        \begin{aligned}
            &-0.5 (1-\cos(\pi t)), &t < 1, \\
            &-1, &t \ge 1.
        \end{aligned}
    \right.
    \nonumber
\end{equation}

The boundaries are all set as slip walls and the particles on free surface are detected and set with $p=0$ at each iteration. The reference density is $\rho_0 = 1$, the gravity force is $G = -1$ in y-direction and sound speed is set as $C_0 = 15$, the dynamics viscosity of $\mu = 2.5\times10^{-3}$ gives a Reynolds number $Re = 400$.

\begin{figure}[!htbp]
	\centering
	\subfigure{\includegraphics[width = 0.8\columnwidth, clip]{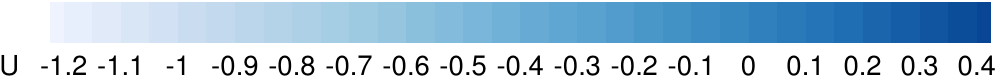}} \\
	\subfigure{\includegraphics[width = 0.48\columnwidth, trim = 30 5 60 30, clip]{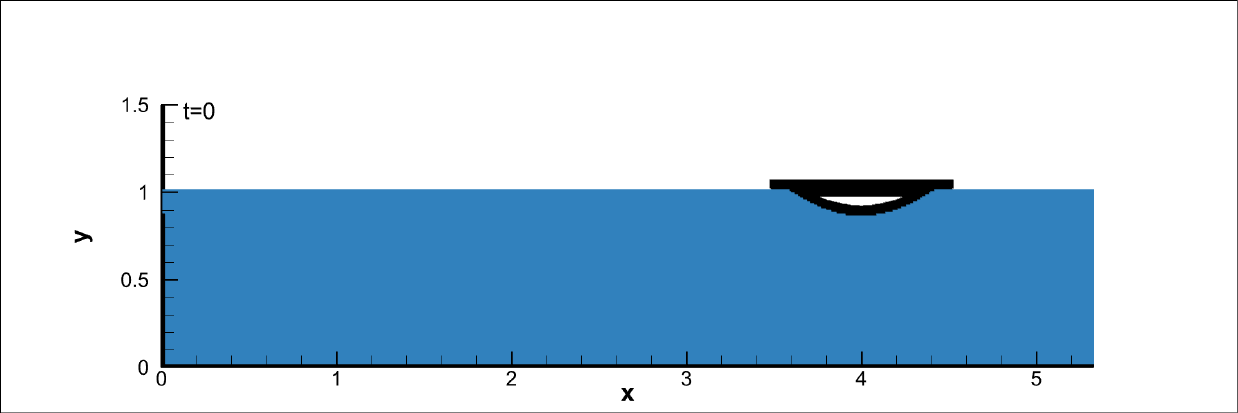}} \hspace{5pt}
	\subfigure{\includegraphics[width = 0.48\columnwidth, trim = 30 5 60 30, clip]{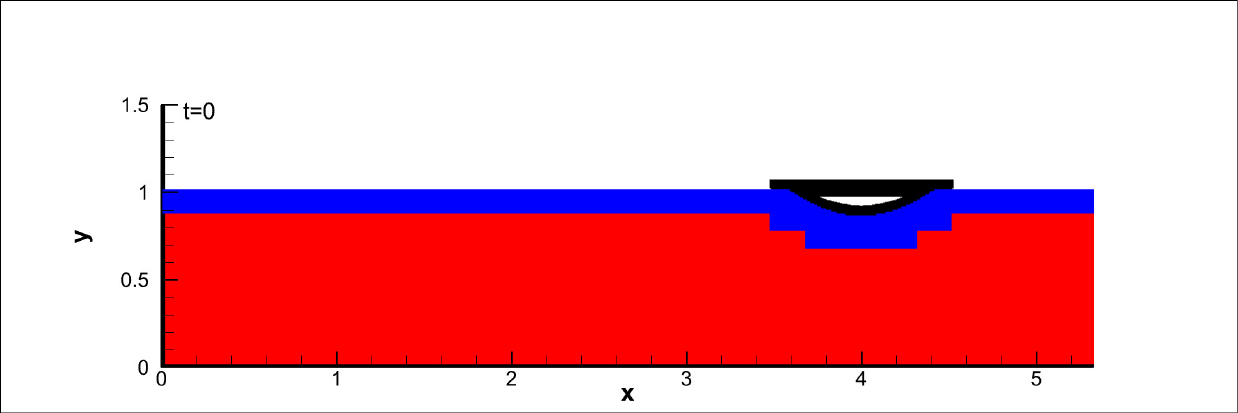}} \\
    \subfigure{\includegraphics[width = 0.48\columnwidth, trim = 30 5 60 30, clip]{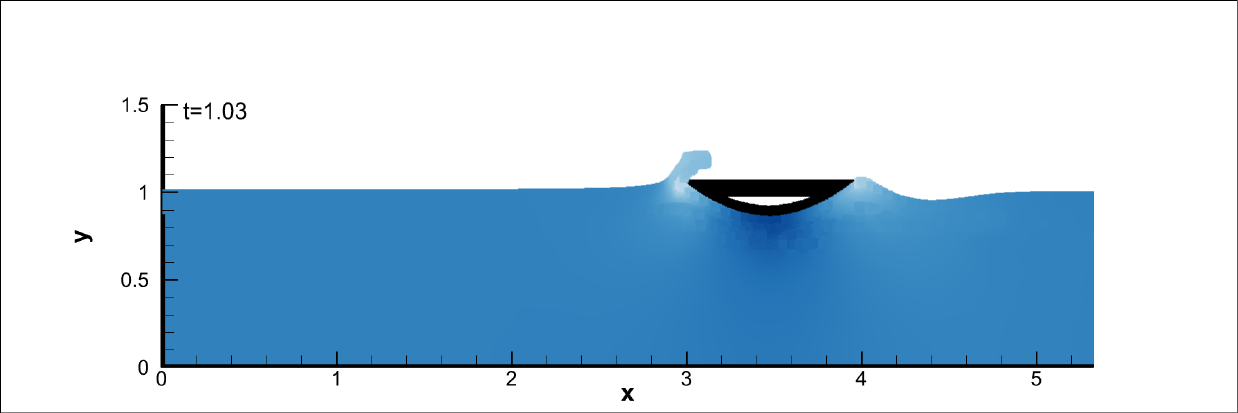}} \hspace{5pt}
	\subfigure{\includegraphics[width = 0.48\columnwidth, trim = 30 5 60 30, clip]{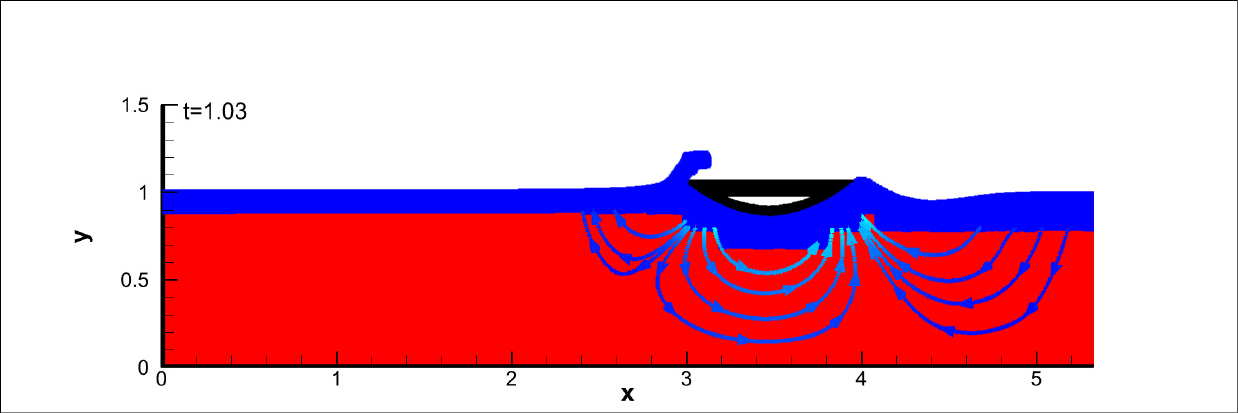}} \\
	\subfigure{\includegraphics[width = 0.48\columnwidth, trim = 30 5 60 30, clip]{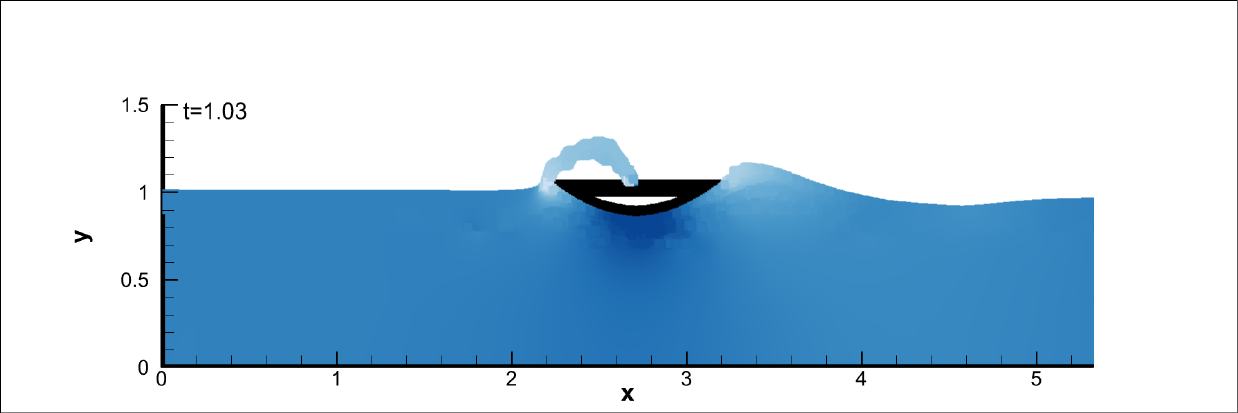}} \hspace{5pt}
	\subfigure{\includegraphics[width = 0.48\columnwidth, trim = 30 5 60 30, clip]{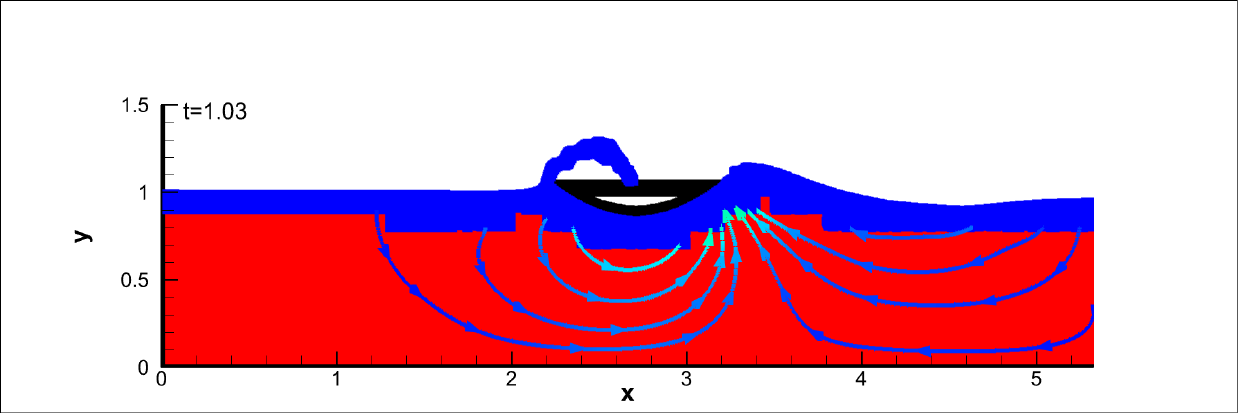}} \\
    \subfigure{\includegraphics[width = 0.48\columnwidth, trim = 30 5 60 30, clip]{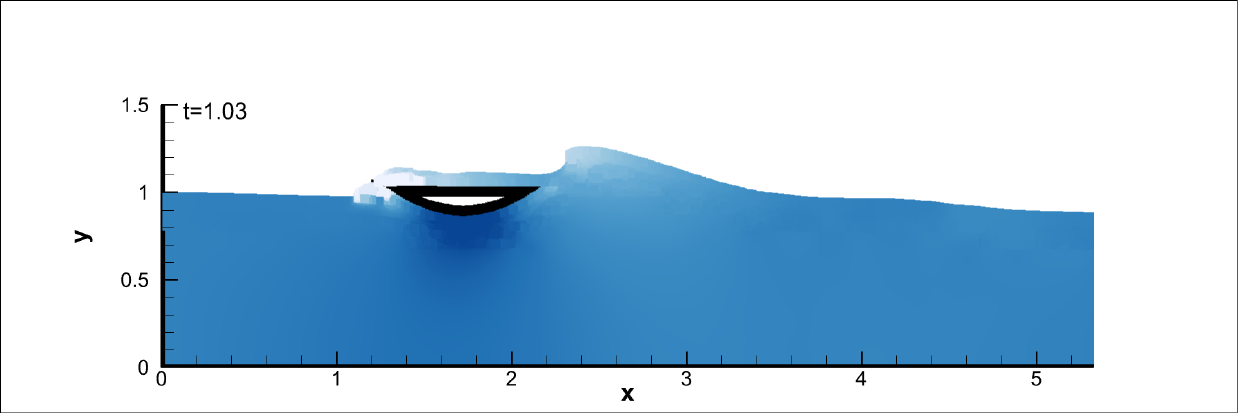}} \hspace{5pt}
	\subfigure{\includegraphics[width = 0.48\columnwidth, trim = 30 5 60 30, clip]{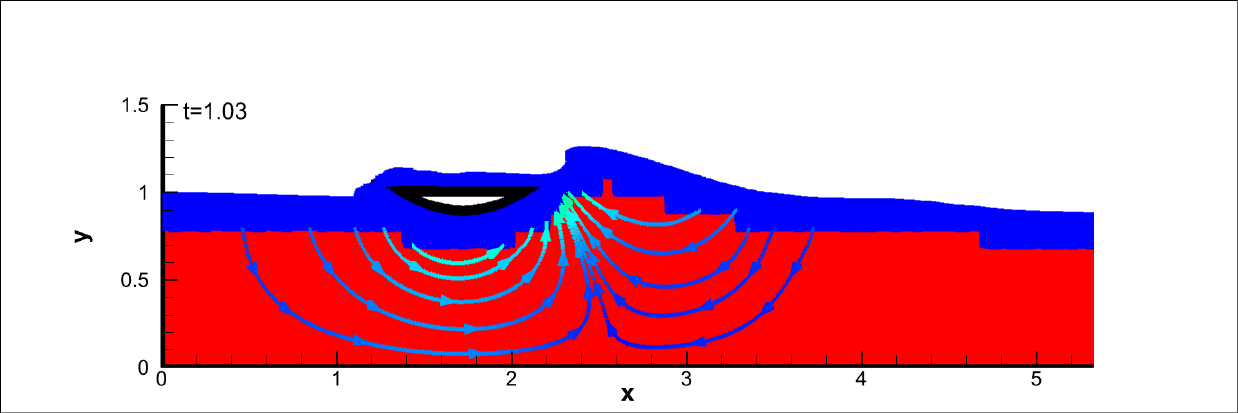}} \\
	\caption{Numerical results of ship cruising problem at $T = 0.0, 1.03, 2.28, 2.78$. Left: $U$-component velocity, right: mesh-particle partitions (red: mesh region, blue: particle region) with streamlines}
	\label{fig:ship}
\end{figure}

\begin{figure}
	\centering
	\includegraphics[width=0.8\linewidth, trim = 2 3 2 2, clip]{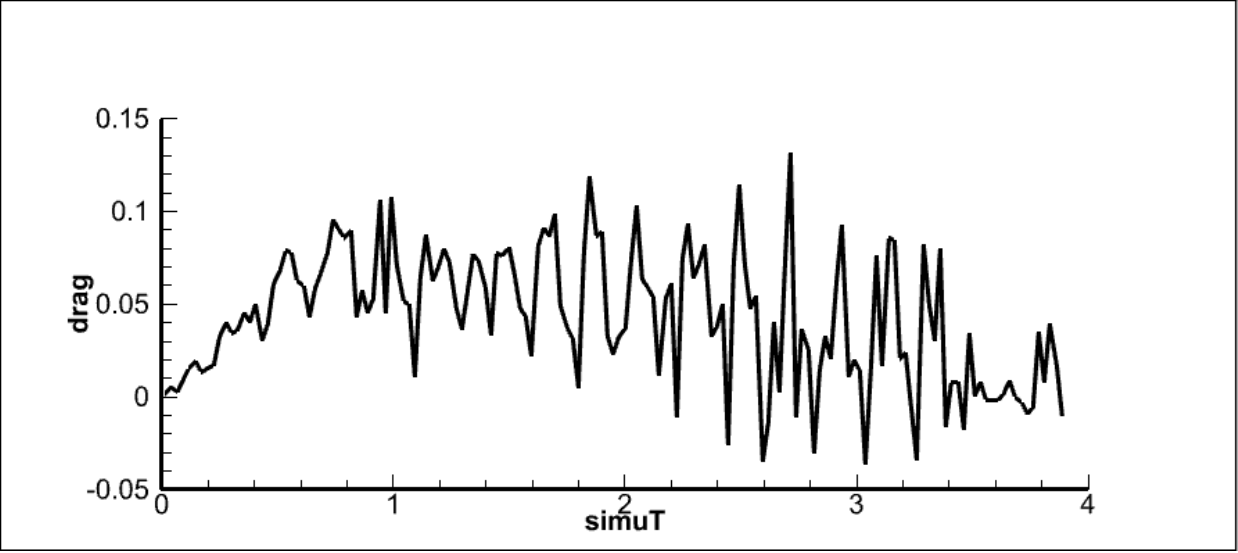}
	\caption{Drag history of ship body during the simulation}
	\label{fig:shipDrag}
\end{figure}

The contours of $U$-component velocity and mesh-particle partitions at $T = 0, 1.03, 2.28, 2.78$ are shown in Figure \ref{fig:ship}, which demonstrate the bow spraying, spray reattachment and the stern wave phenomena clearly, and the right column shows the mesh and particle partition with streamlines colored by the magnitude of velocity, where the mesh blocks and particles block convert to each other dynamically and adaptively with the ship cruises. The drag history of the ship body is demonstrated in Figure \ref{fig:shipDrag}. 

\subsection{Body entry problem}
\begin{figure}
    \centering
    \includegraphics[width=0.5\linewidth, clip]{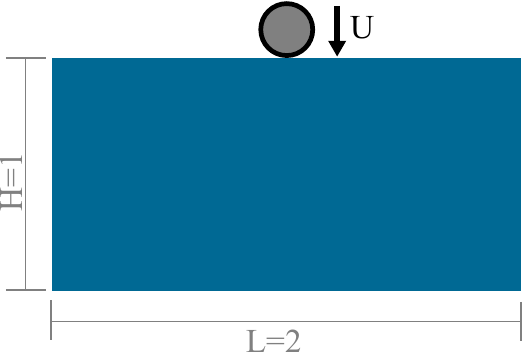}
    \caption{Schematic for cylinder-body water entry problem}
    \label{fig:bodyEntry}
\end{figure}

The cylinder-body water entry problem is investigated by AFVPM. The computational domain is shown in Figure \ref{fig:bodyEntry}, where a cylinder of diameter $D = 0.11$ falls into a tank of $L=2,H=1$ at a constant speed $V = -1.5$.

The initial particle space and mesh size is $\Delta x = 1/400$, thus a total of 320000 particles / mesh cells are generated initially. The boundaries are all set as slip walls. The reference density is $\rho_0 = 1$, the gravity force is $G = -1$ in y-direction and sound speed is set as $C_0 = 22.5$, the dynamics viscosity of $\mu = 1.65\times10^{-3}$ gives a Reynolds number $Re = 100$.

\begin{figure}[!htbp]
	\centering
    \subfigure{\includegraphics[width = 0.8\columnwidth, clip]{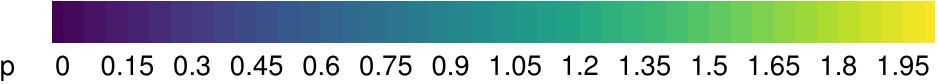}} \\
	\subfigure{\includegraphics[width = 0.48\columnwidth, trim = 30 5 50 30, clip]{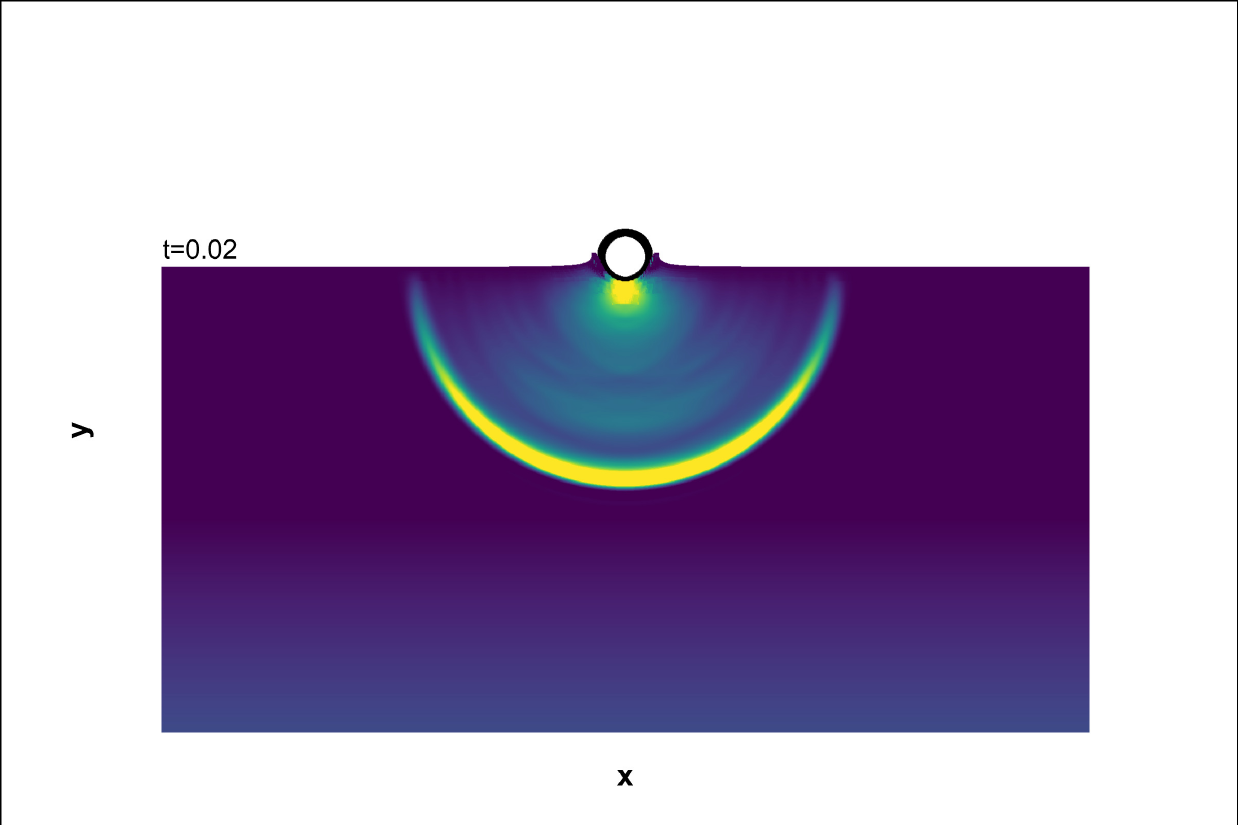}} \hspace{5pt}
	\subfigure{\includegraphics[width = 0.48\columnwidth, trim = 30 5 50 30, clip]{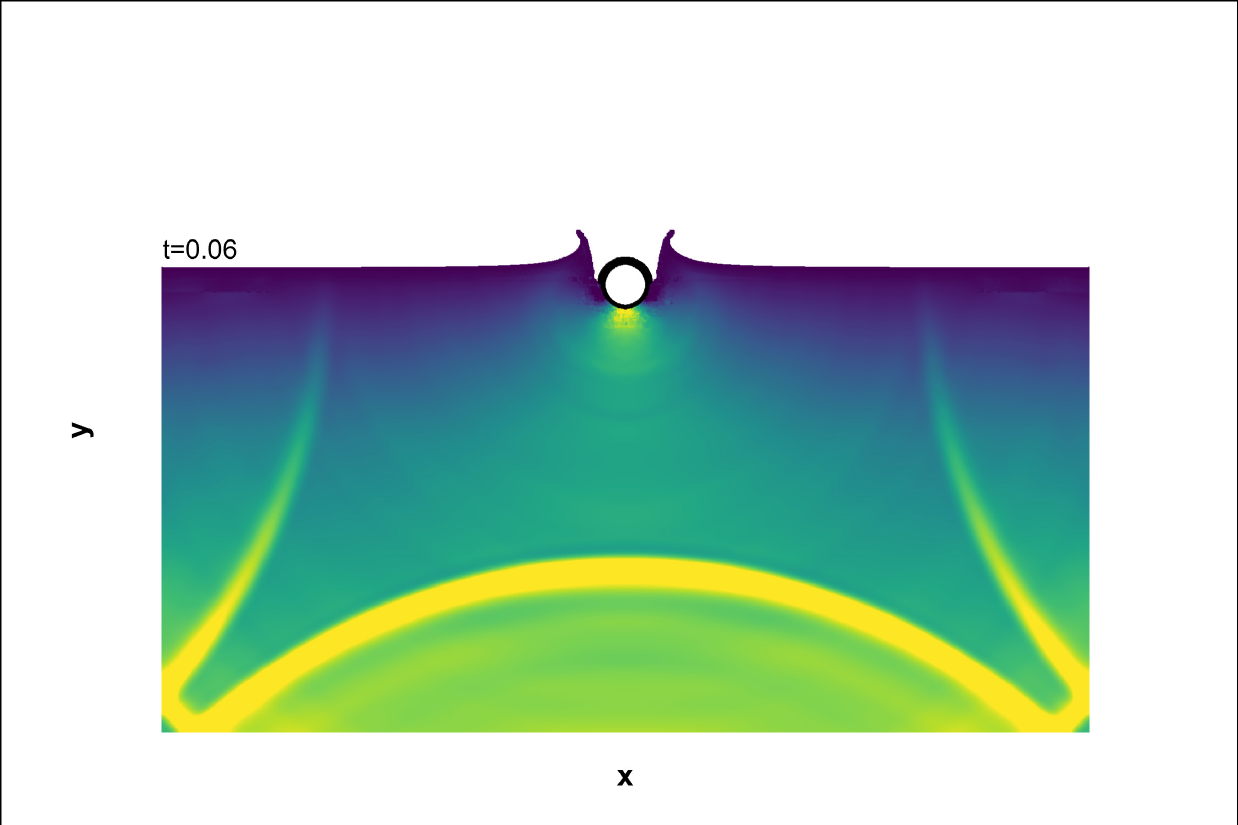}} \\
	\subfigure{\includegraphics[width = 0.48\columnwidth, trim = 30 5 50 30, clip]{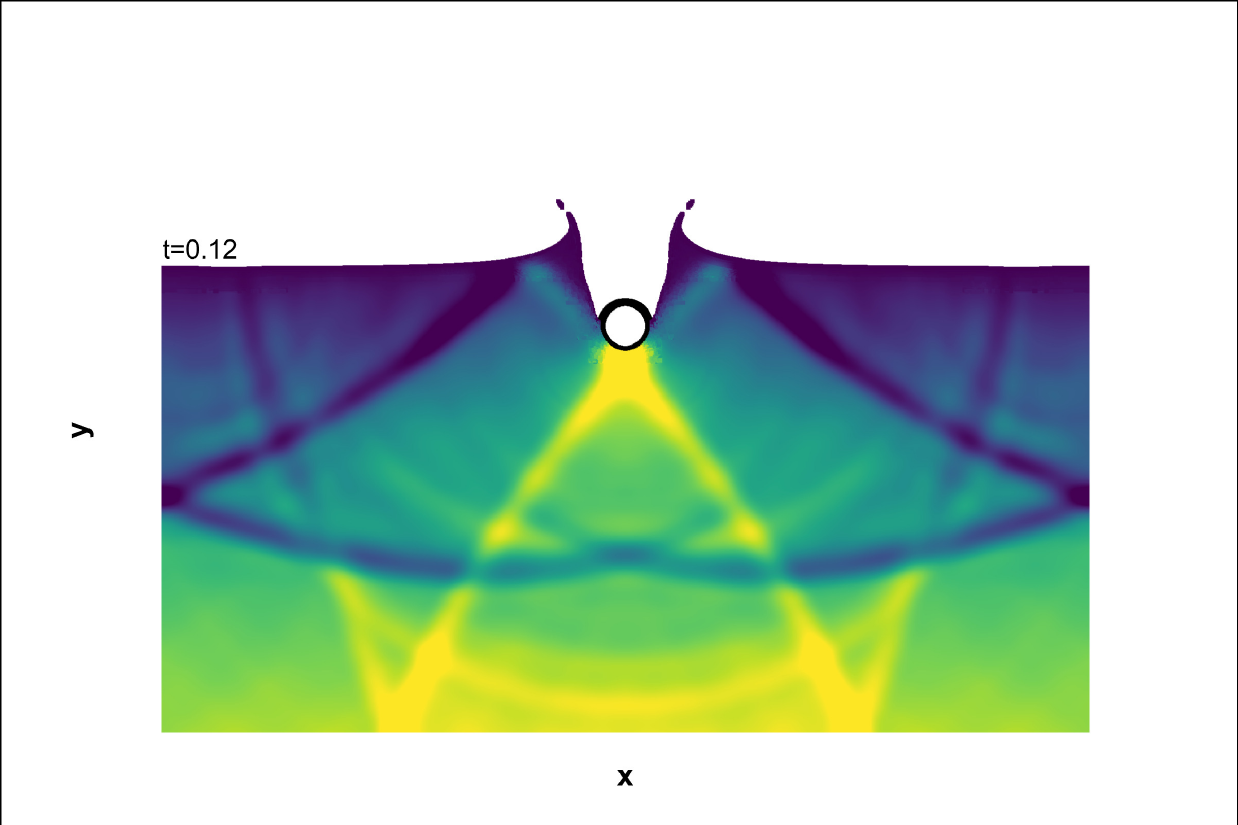}} \hspace{5pt}
	\subfigure{\includegraphics[width = 0.48\columnwidth, trim = 30 5 50 30, clip]{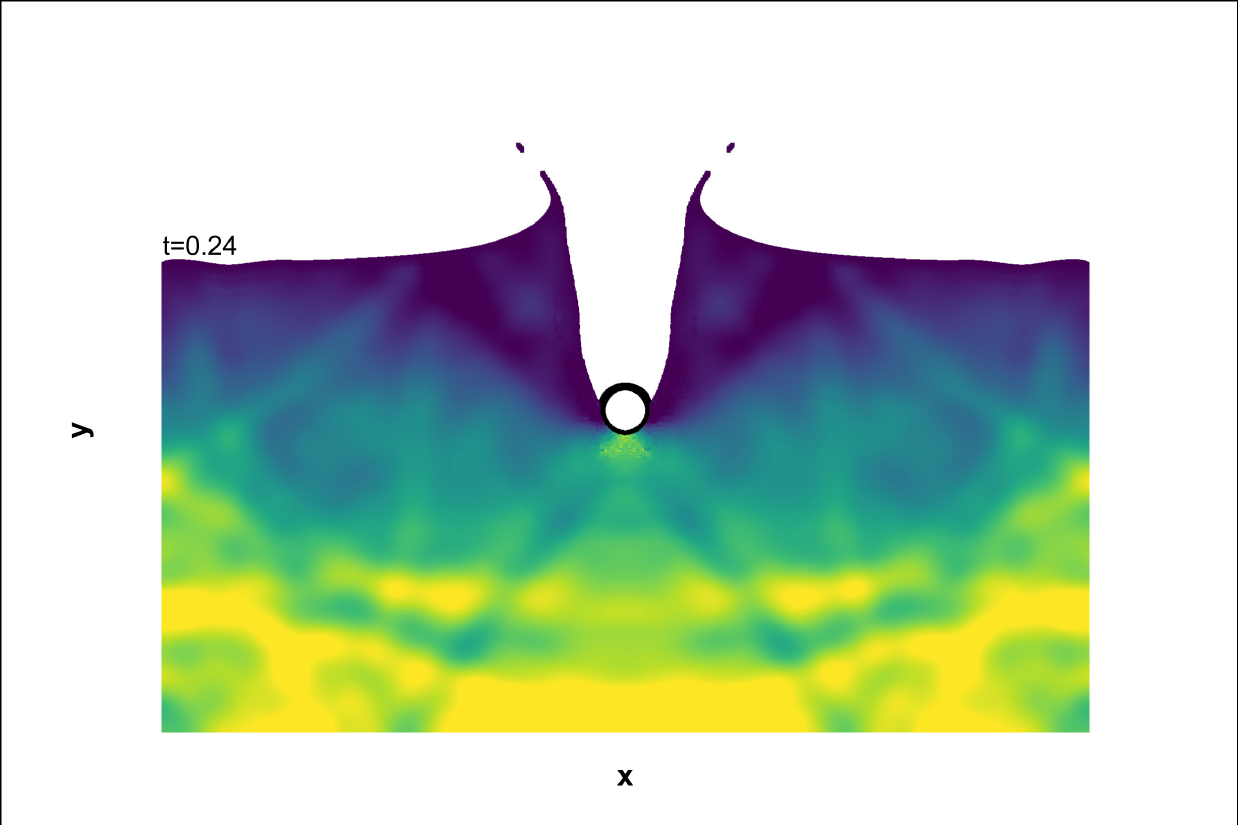}} \\
	\caption{Propagation of pressure wave of body entry problem at $T = 0.02, 0.06, 0.12, 0.24$}
	\label{fig:entry-press}
\end{figure}

\begin{figure}[!htbp]
	\centering
	\subfigure{\includegraphics[width = 0.48\columnwidth, trim = 30 10 200 70, clip]{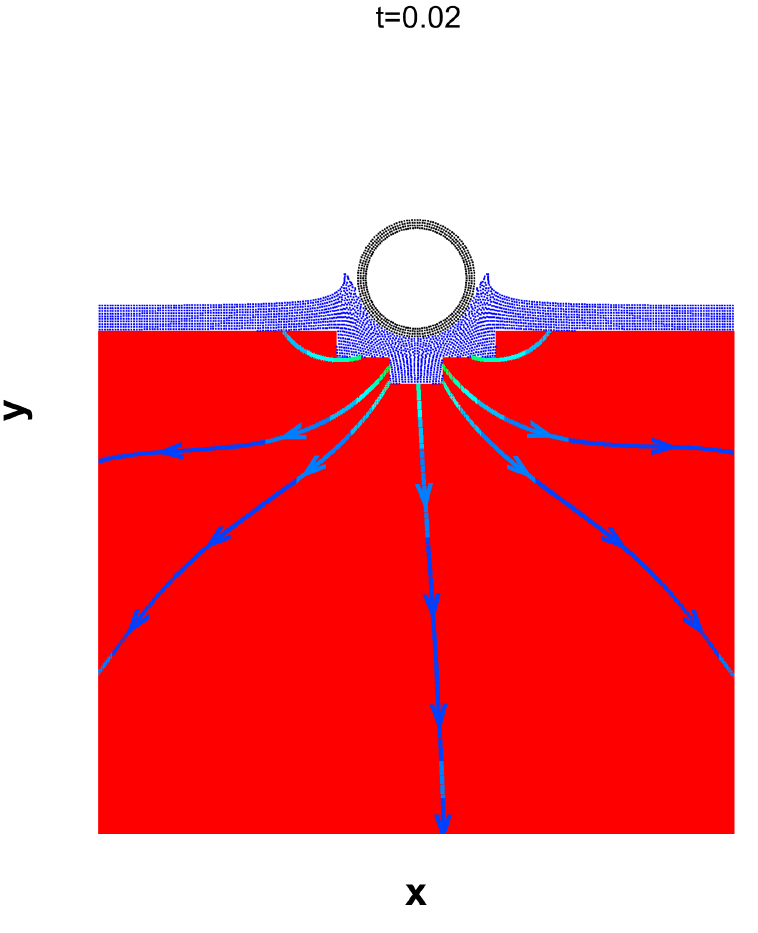}} \hspace{5pt}
	\subfigure{\includegraphics[width = 0.48\columnwidth, trim = 30 10 200 70, clip]{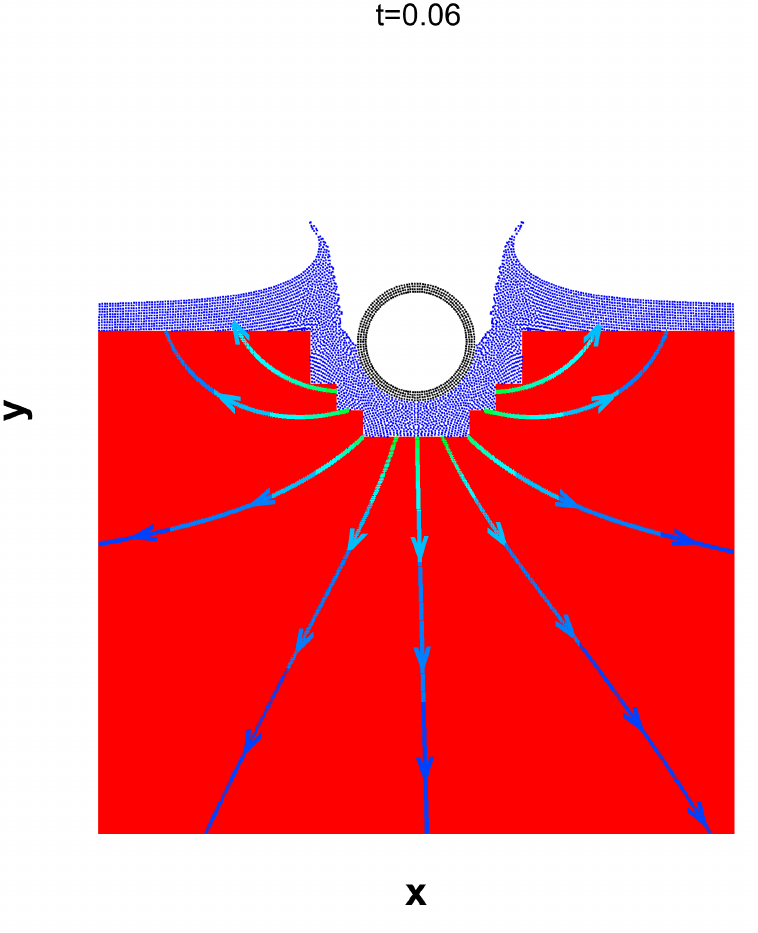}} \\
	\subfigure{\includegraphics[width = 0.48\columnwidth, trim = 30 10 200 70, clip]{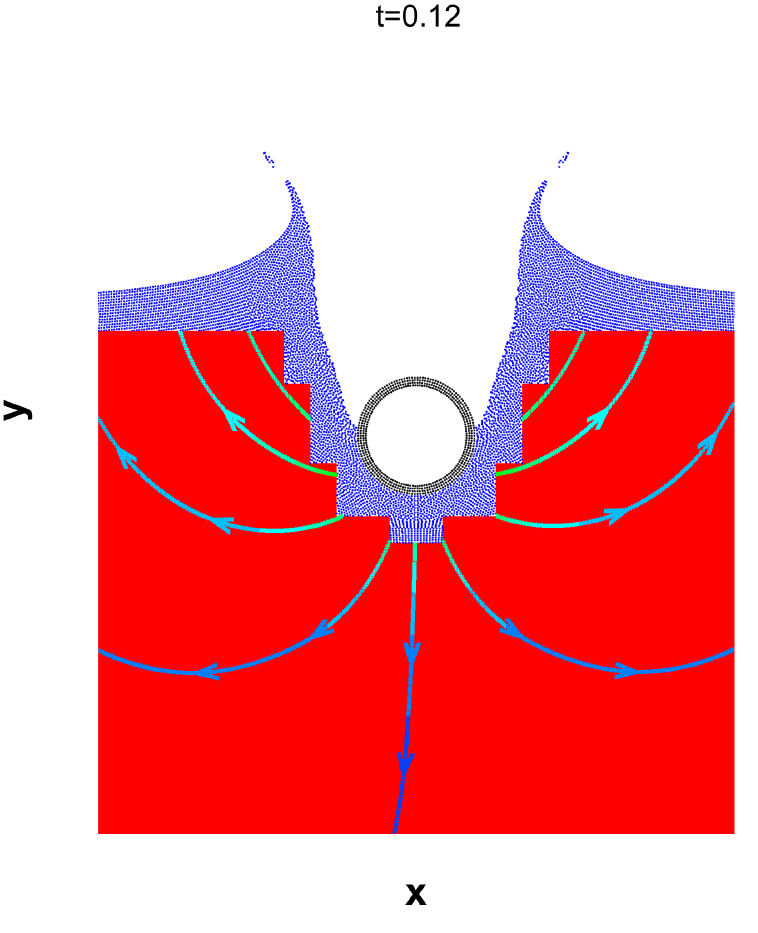}} \hspace{5pt}
	\subfigure{\includegraphics[width = 0.48\columnwidth, trim = 30 10 200 70, clip]{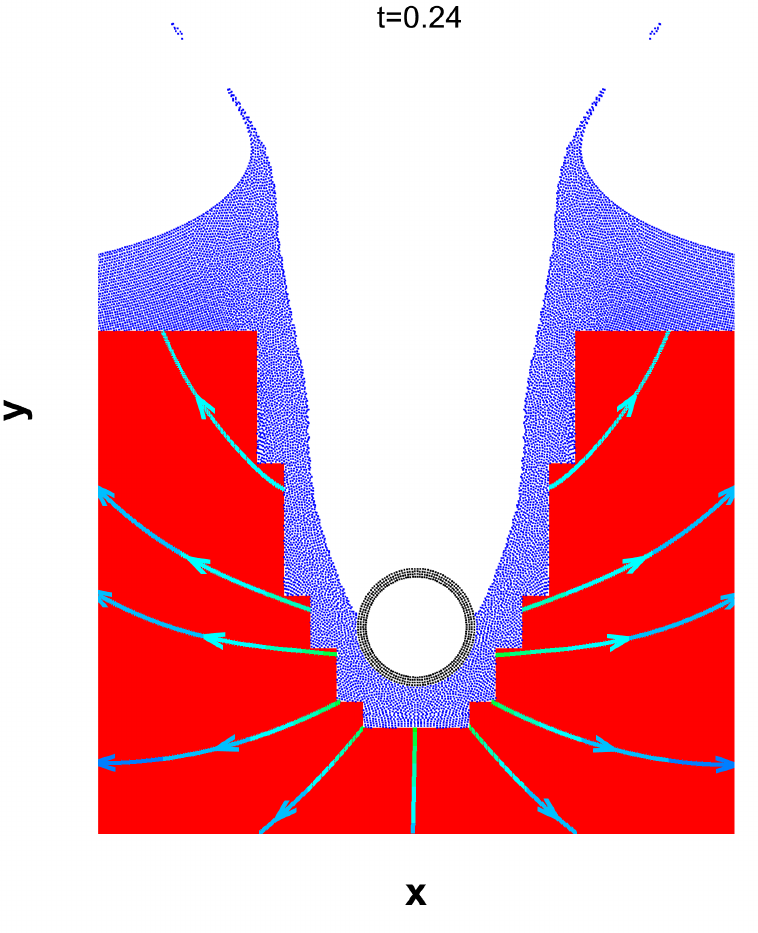}} \\
	\caption{Mesh-particle partitions (red: mesh region, blue: particle region) with streamlines of body entry problem at $T = 0.02, 0.06, 0.12, 0.24$}
	\label{fig:entry-partition}
\end{figure}

\begin{figure}
	\centering
	\includegraphics[width=0.6\linewidth, trim = 10 10 10 10, clip]{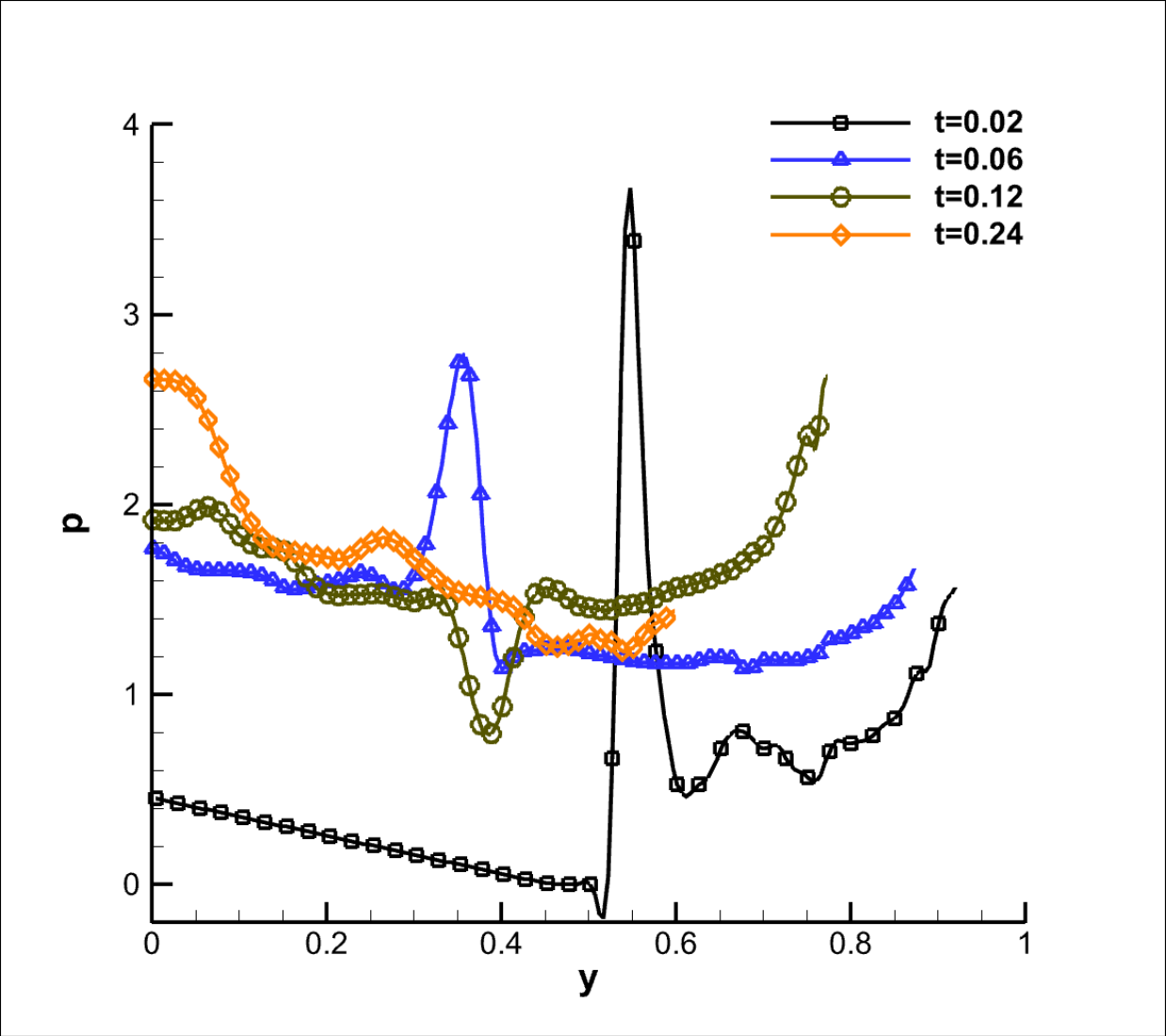}
	\caption{Pressure distribution along the vertical line $x=1$ from water bottom to cylinder surface of body entry problem}
	\label{fig:entryPress}
\end{figure}

The propagation of pressure wave is shown in Figure \ref{fig:entry-press}. The pressure wave spreads once the cylinder hits the water surface, collides with the tank walls, reflects and dissipates gradually. The dynamics conversion of mesh and particle regions are shown in Figure \ref{fig:entry-partition}, where mesh and particle blocks convert to each other dynamically and adaptively, keeping uniform particle distribution near the free surface and smooth transition across mesh and particle interfaces, the streamlines colored by the magnitude of velocity are also shown in Figure \ref{fig:entry-partition} to show the flows near the cylinder. The pressure distributions along $x=1$ from the water bottom to the cylinder at different computational times are shown in Figure \ref{fig:entryPress}.

\begin{figure}[!htbp]
	\centering
	\subfigure[full SPH]{\includegraphics[width = 0.48\columnwidth, trim = 30 10 150 70, clip]{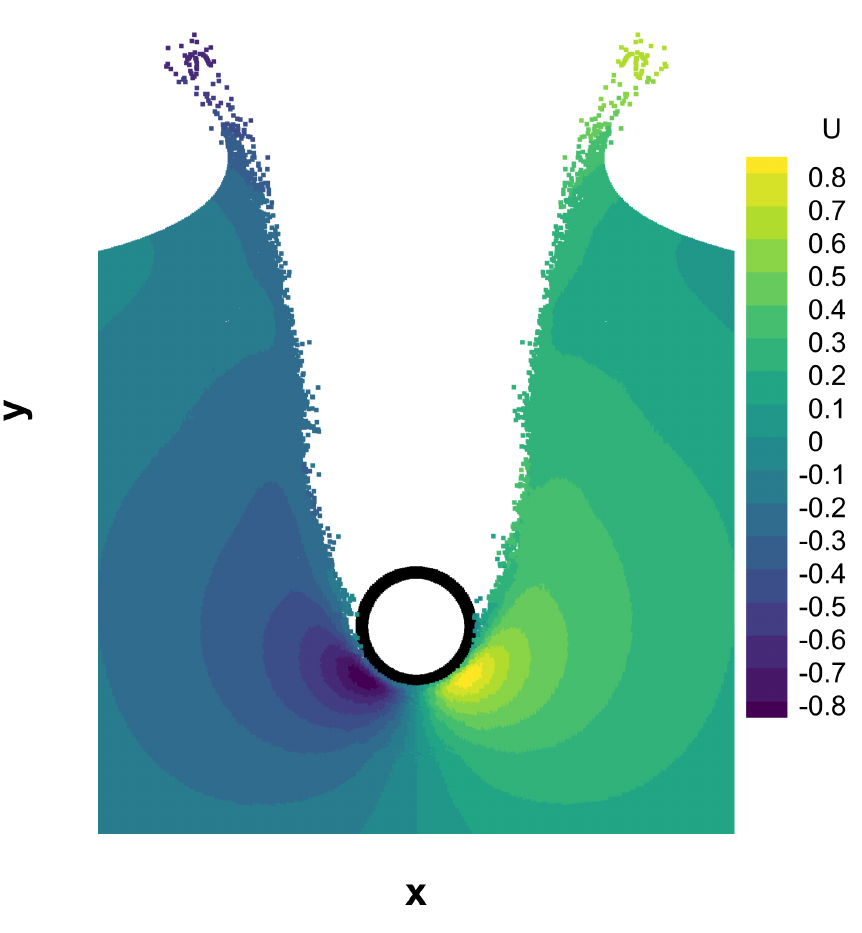}} \hspace{5pt}
	\subfigure[AFVPM]{\includegraphics[width = 0.48\columnwidth, trim = 30 10 150 70, clip]{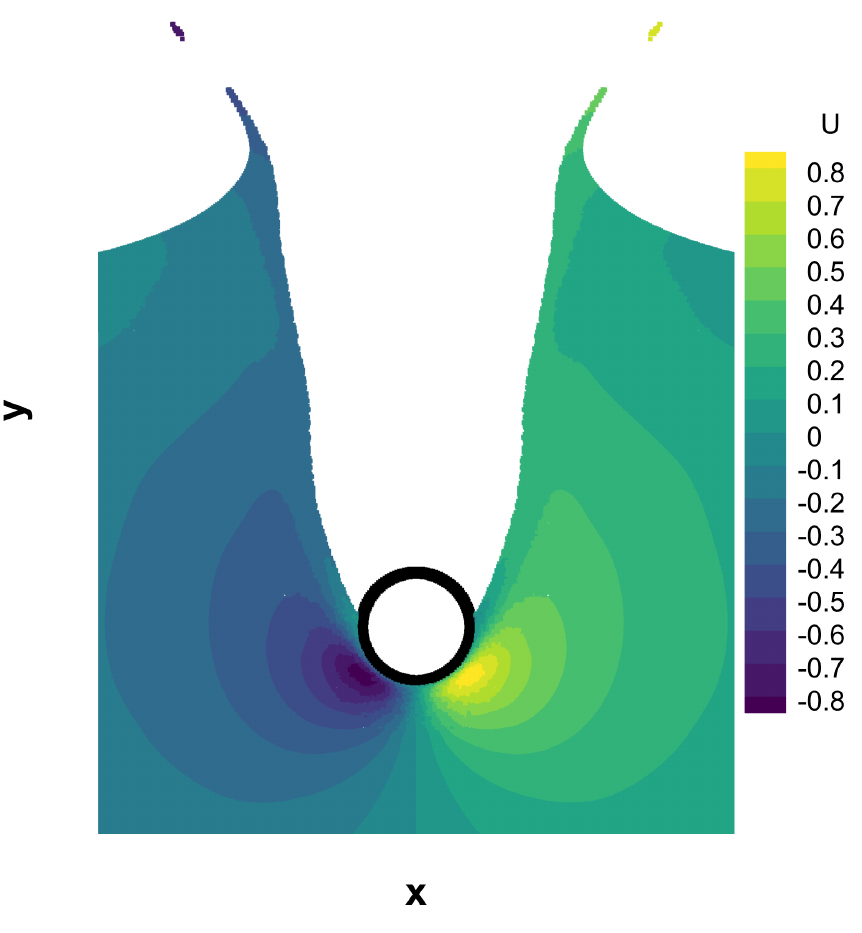}} \\
	\caption{Comparison between AFVPM and full SPH on body entry problem}
	\label{fig:entryComp}
\end{figure}

The contour of $U$-component velocity and particle distribution around the cylinder at $T = 0.24$ obtained by full SPH and AFVPM are compared in Figure \ref{fig:entryComp}. AFVPM demonstrate better free-surface resolution and better symmetry compared with the full SPH approach.

\subsection{Water filling problem}
\begin{figure}
    \centering
    \includegraphics[width=0.4\linewidth, clip]{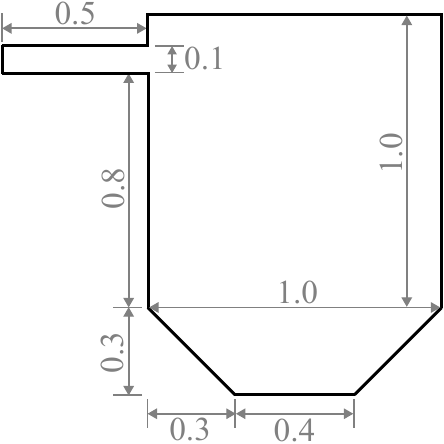}
    \caption{Schematic for water filling problem}
    \label{fig:waterFilling}
\end{figure}

The problem of filling water into a container involves water injection, impacting on walls and water filling processes, and the concurrence of these phenomena pose a great challenge to AFVPM. The schematic of this problem is demonstrated in Figure \ref{fig:waterFilling}, where water is injected into the container from the upper inlet at a uniform speed $U = 0.5$.

The initial particle space and mesh size is $\Delta x = 1/100$. The boundaries are all set as slip walls. The reference density is $\rho_0 = 1$, the gravity force is $G = -1$ in y-direction and sound speed is set as $C_0 = 25$, the dynamics viscosity of $\mu = 4.33\times10^{-3}$ gives a Reynolds number $Re = 400$.

\begin{figure}[!htbp]
	\centering
	\subfigure{\includegraphics[width = 0.45\columnwidth, trim = 10 10 10 10, clip]{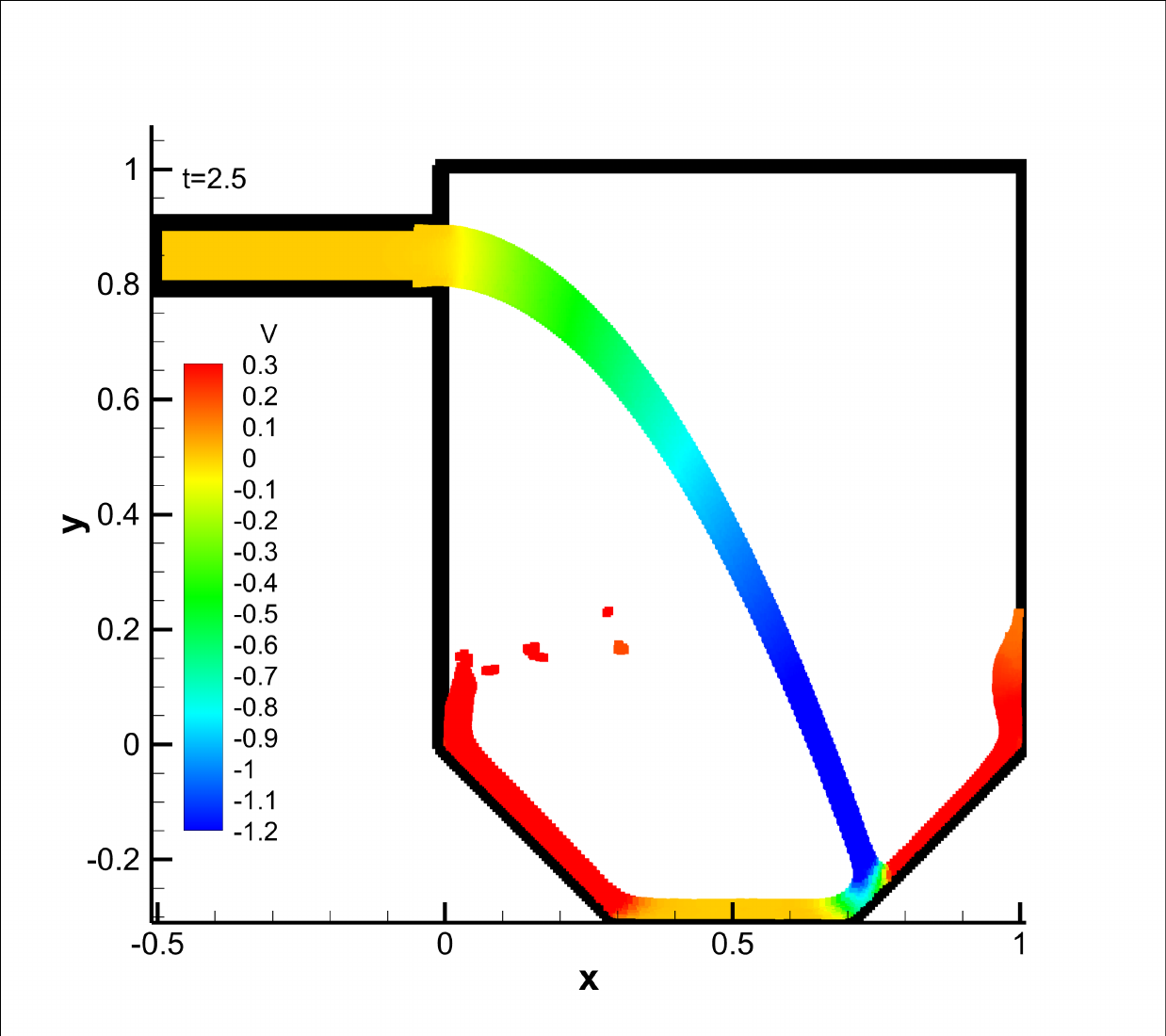}} \hspace{5pt}
	\subfigure{\includegraphics[width = 0.45\columnwidth, trim = 10 10 10 10, clip]{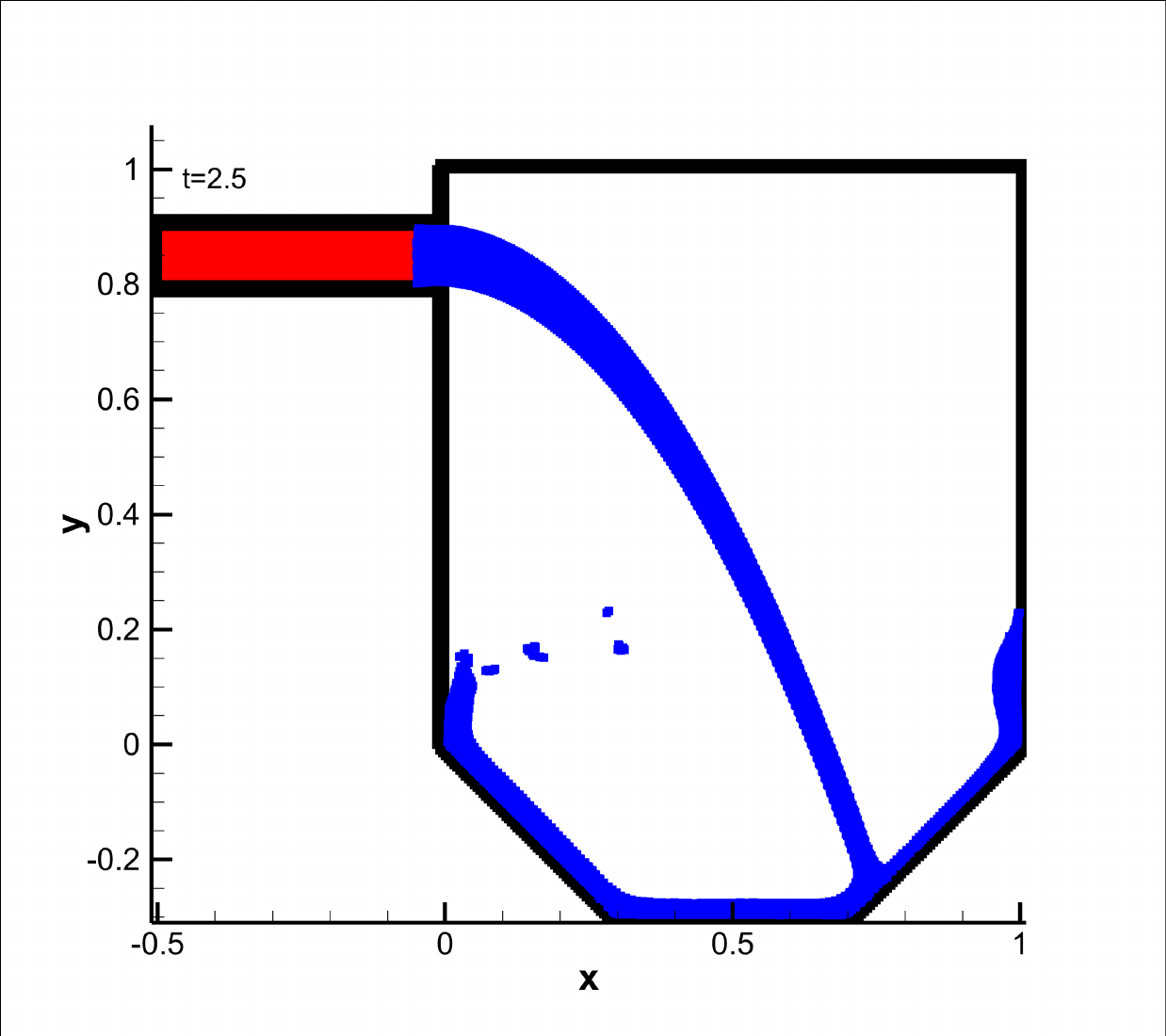}} \\
	\subfigure{\includegraphics[width = 0.45\columnwidth, trim = 10 10 10 10, clip]{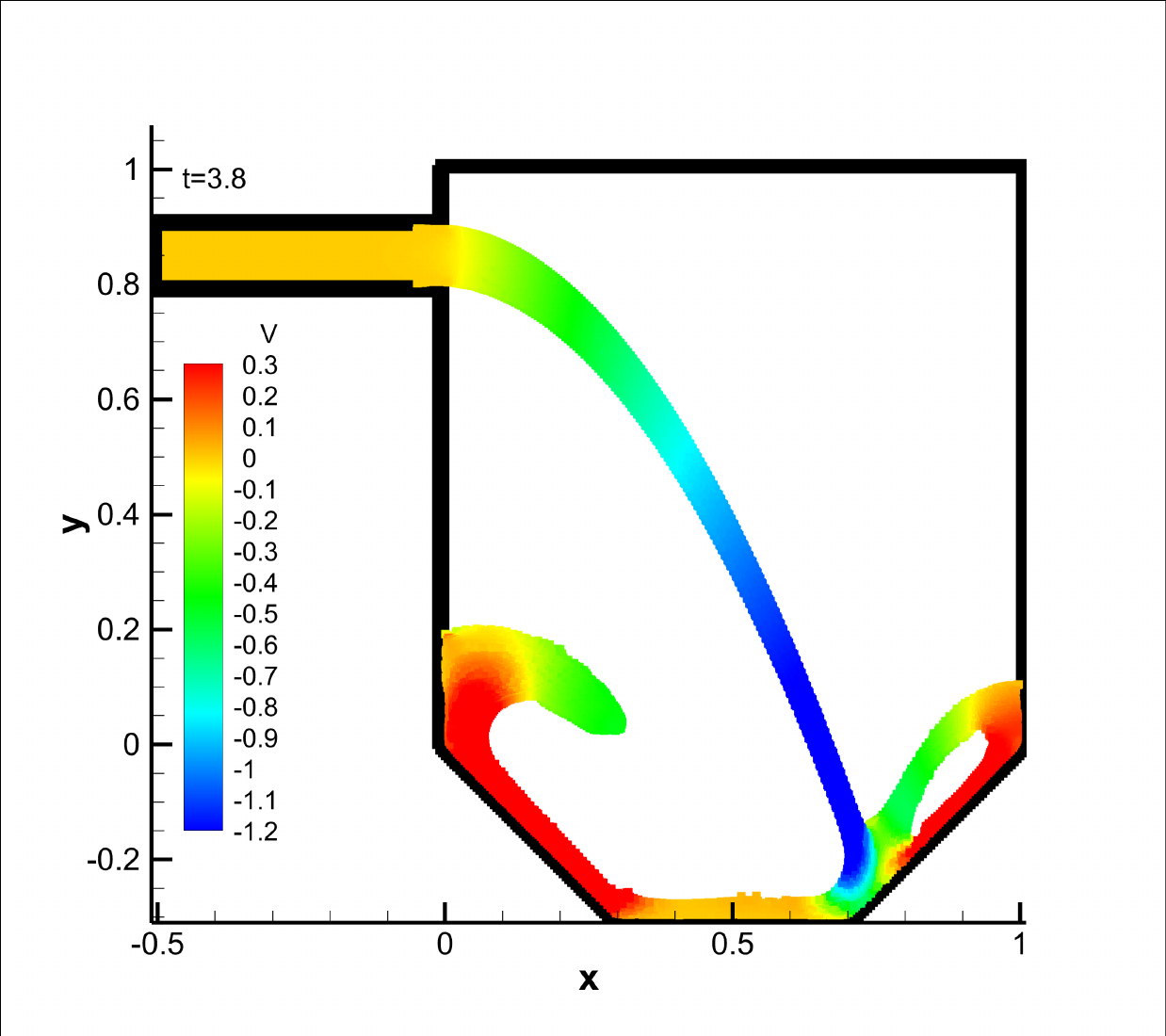}} \hspace{5pt}
	\subfigure{\includegraphics[width = 0.45\columnwidth, trim = 10 10 10 10, clip]{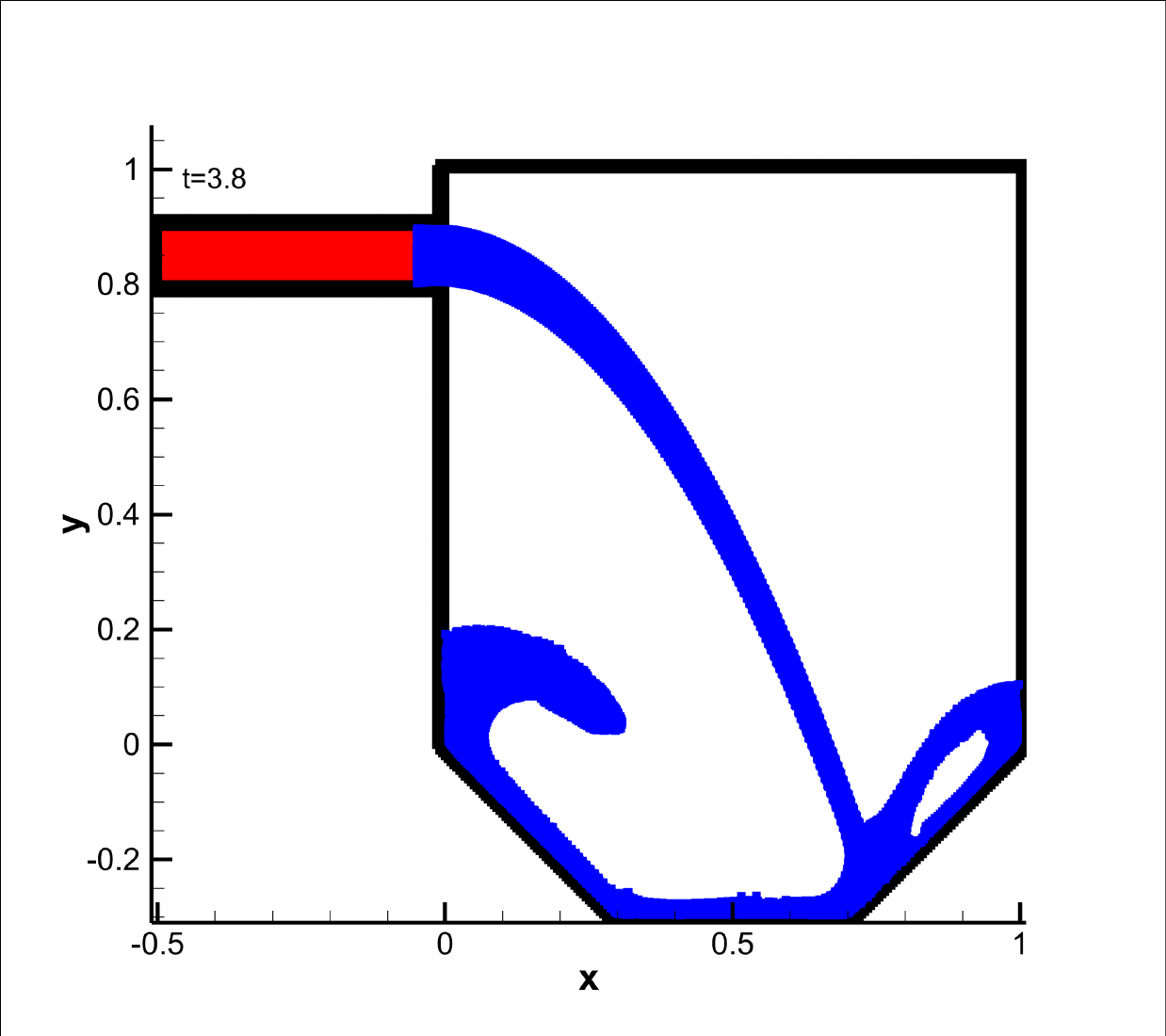}} \\
	\subfigure{\includegraphics[width = 0.45\columnwidth, trim = 10 10 10 10, clip]{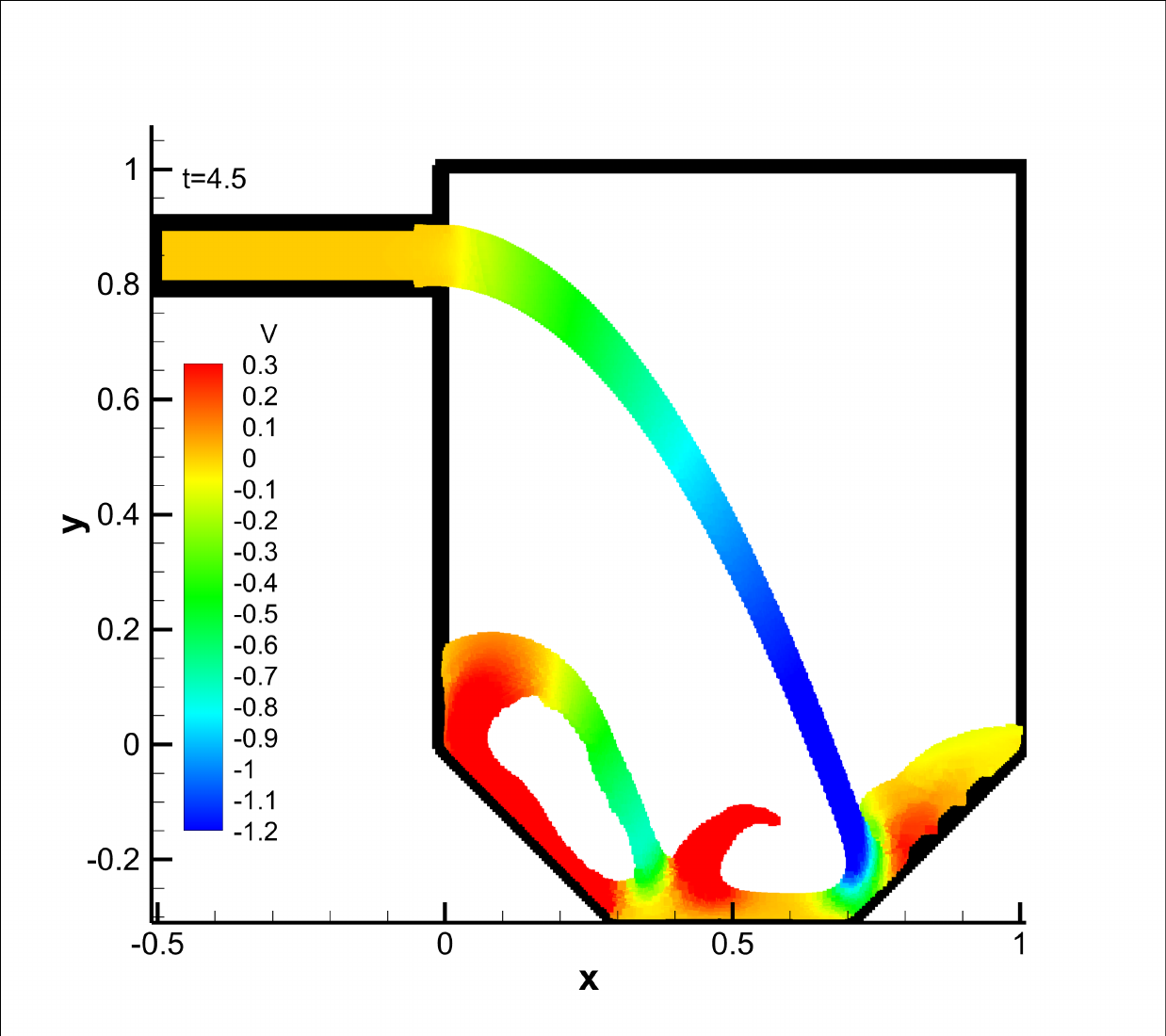}} \hspace{5pt}
	\subfigure{\includegraphics[width = 0.45\columnwidth, trim = 10 10 10 10, clip]{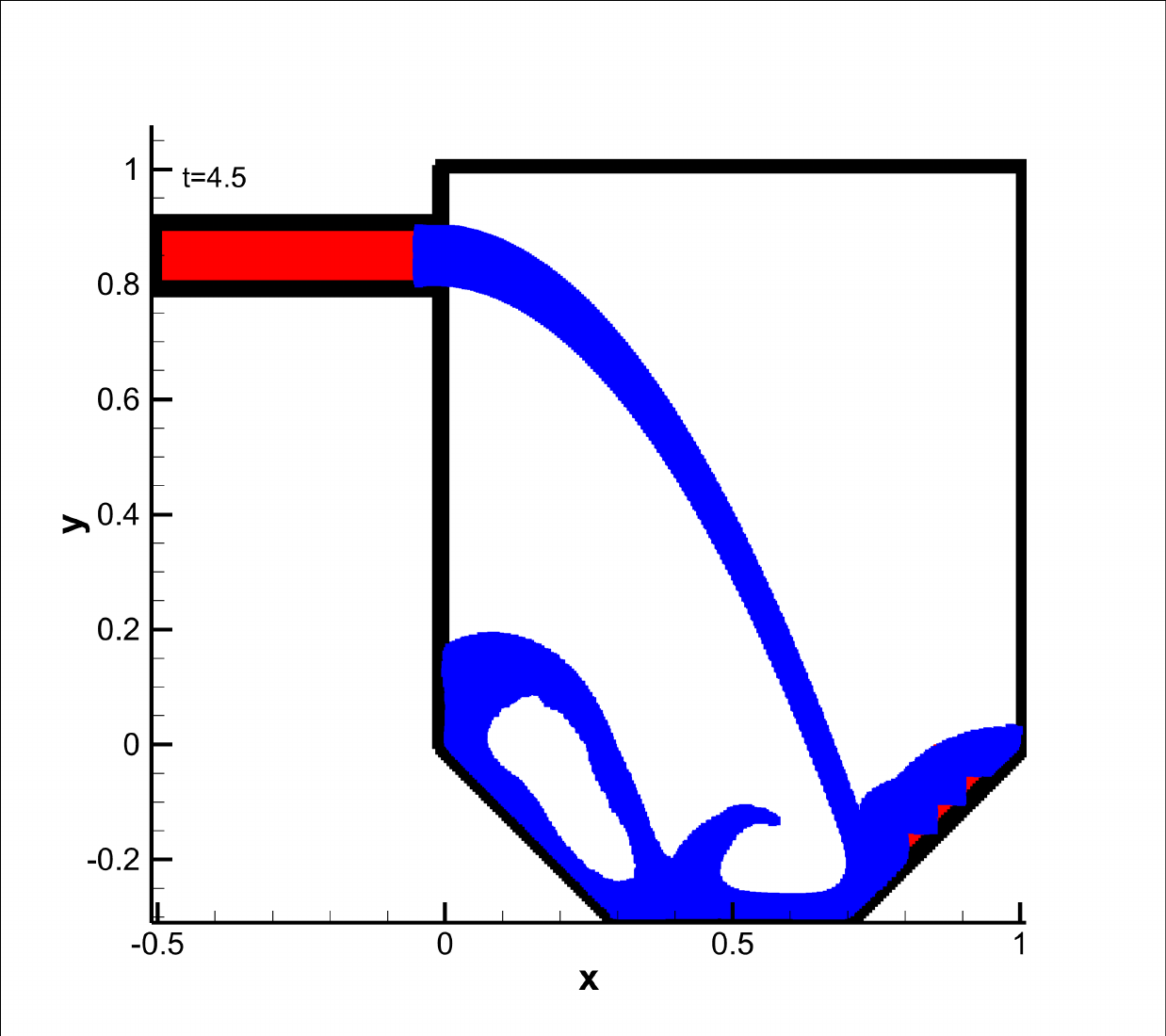}} \\
	\caption{Numerical results of water filling problem at $T = 2.5, 3.8, 4.5$. Left: $V$-component velocity, right: mesh-particle partitions (red: mesh region, blue: particle region)}
	\label{fig:flooding1}
\end{figure}

\begin{figure}[!htbp]
	\centering
	\subfigure{\includegraphics[width = 0.45\columnwidth, trim = 10 10 10 10, clip]{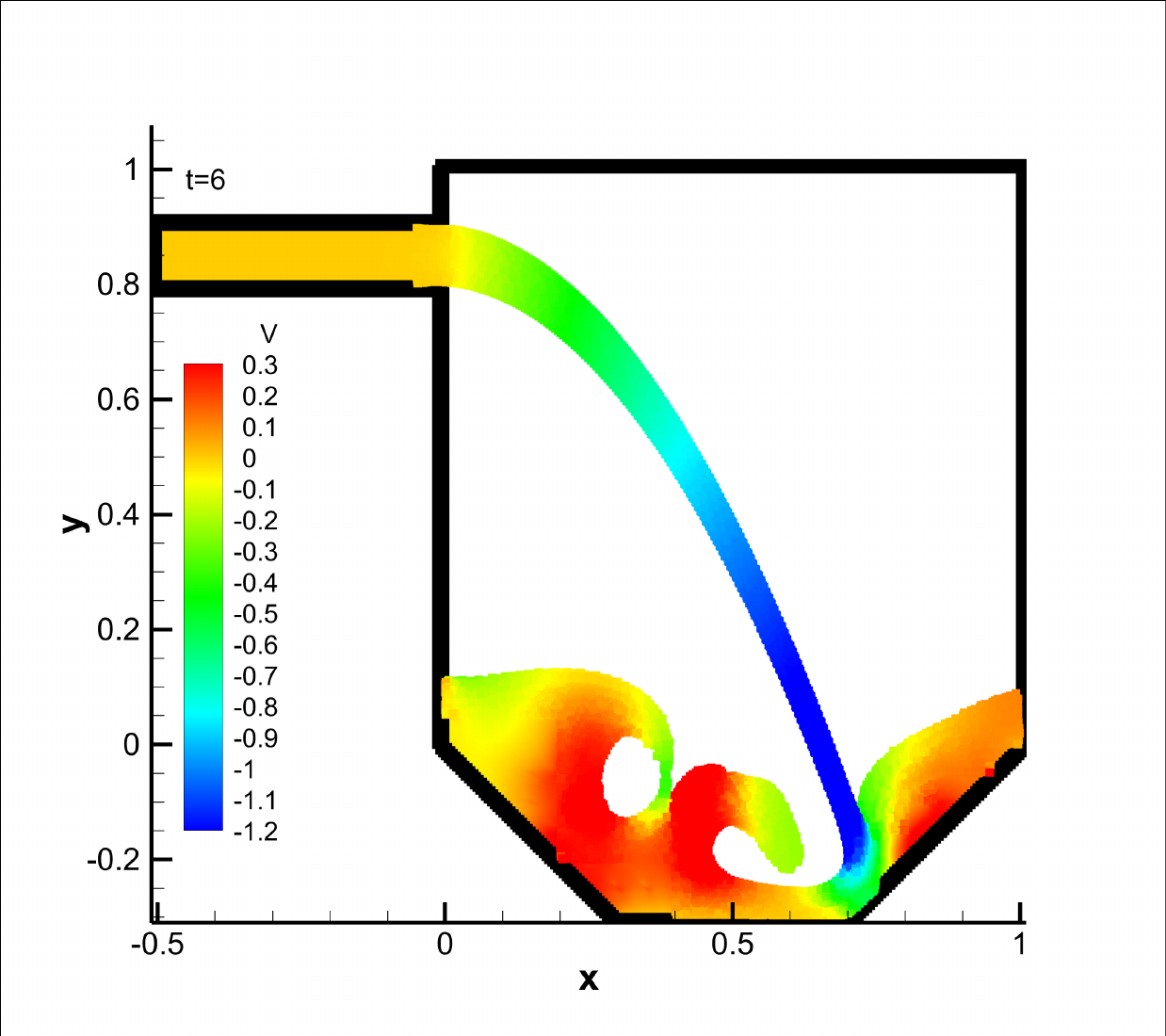}} \hspace{5pt}
	\subfigure{\includegraphics[width = 0.45\columnwidth, trim = 10 10 10 10, clip]{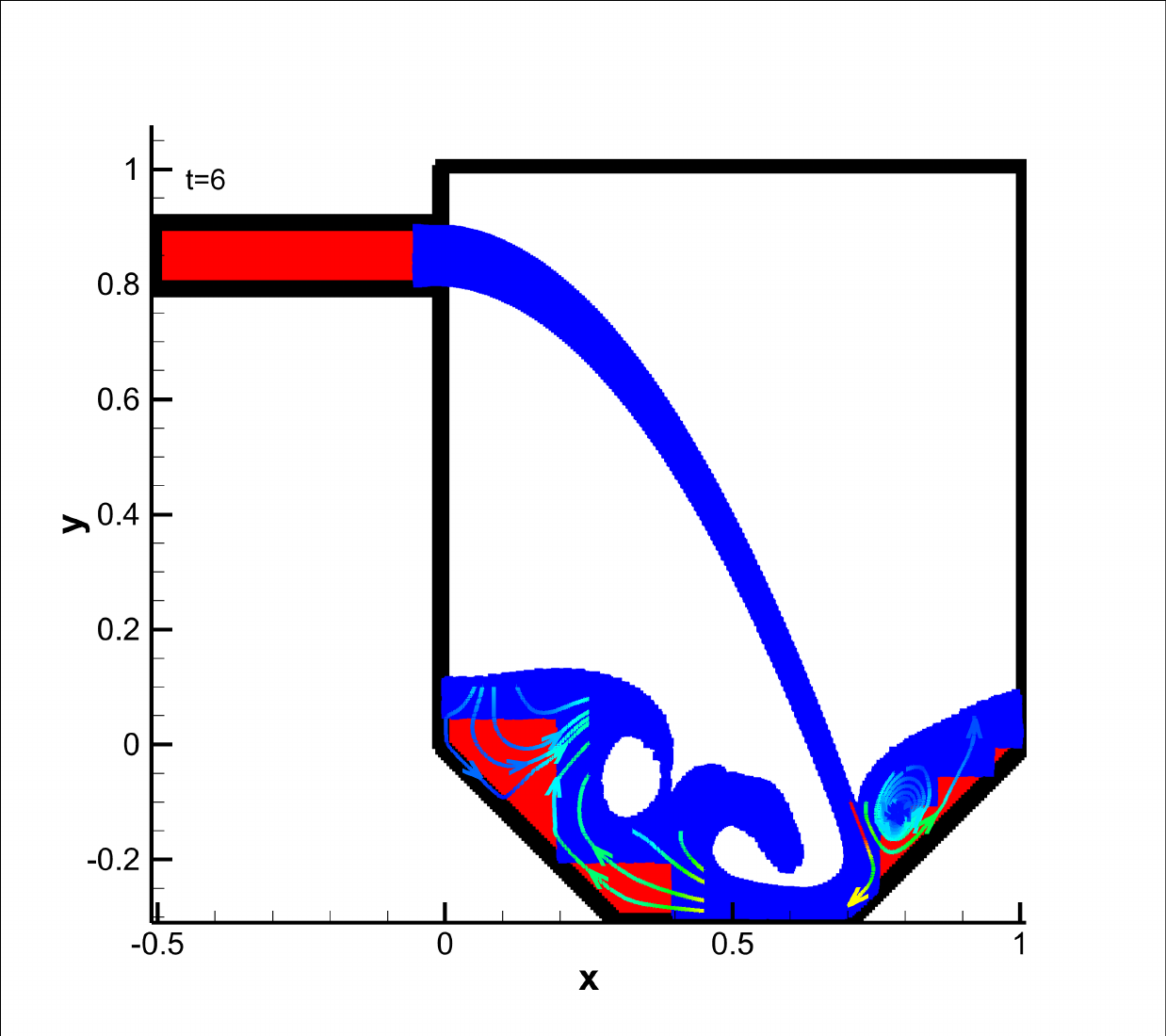}} \\
	\subfigure{\includegraphics[width = 0.45\columnwidth, trim = 10 10 10 10, clip]{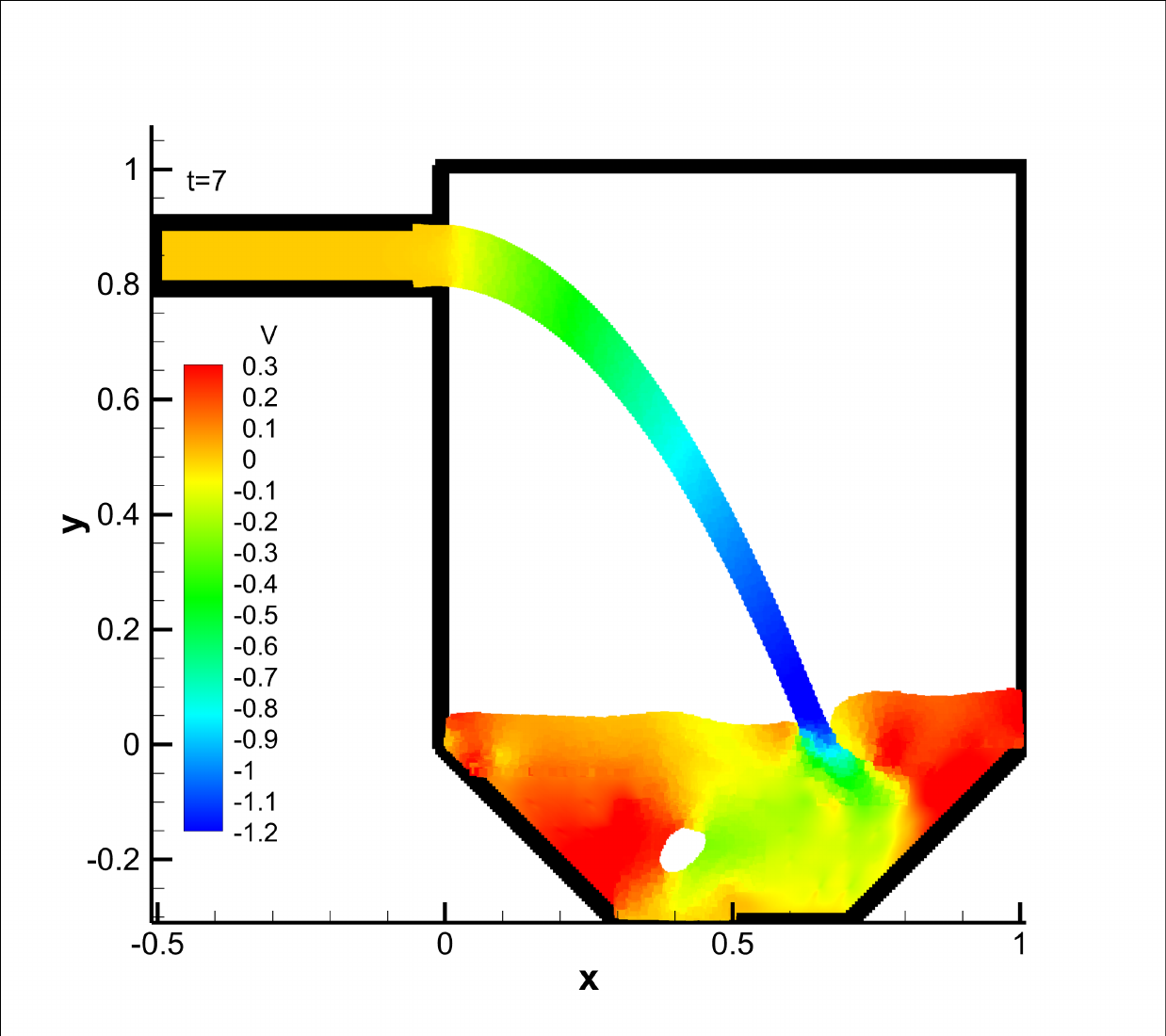}} \hspace{5pt}
	\subfigure{\includegraphics[width = 0.45\columnwidth, trim = 10 10 10 10, clip]{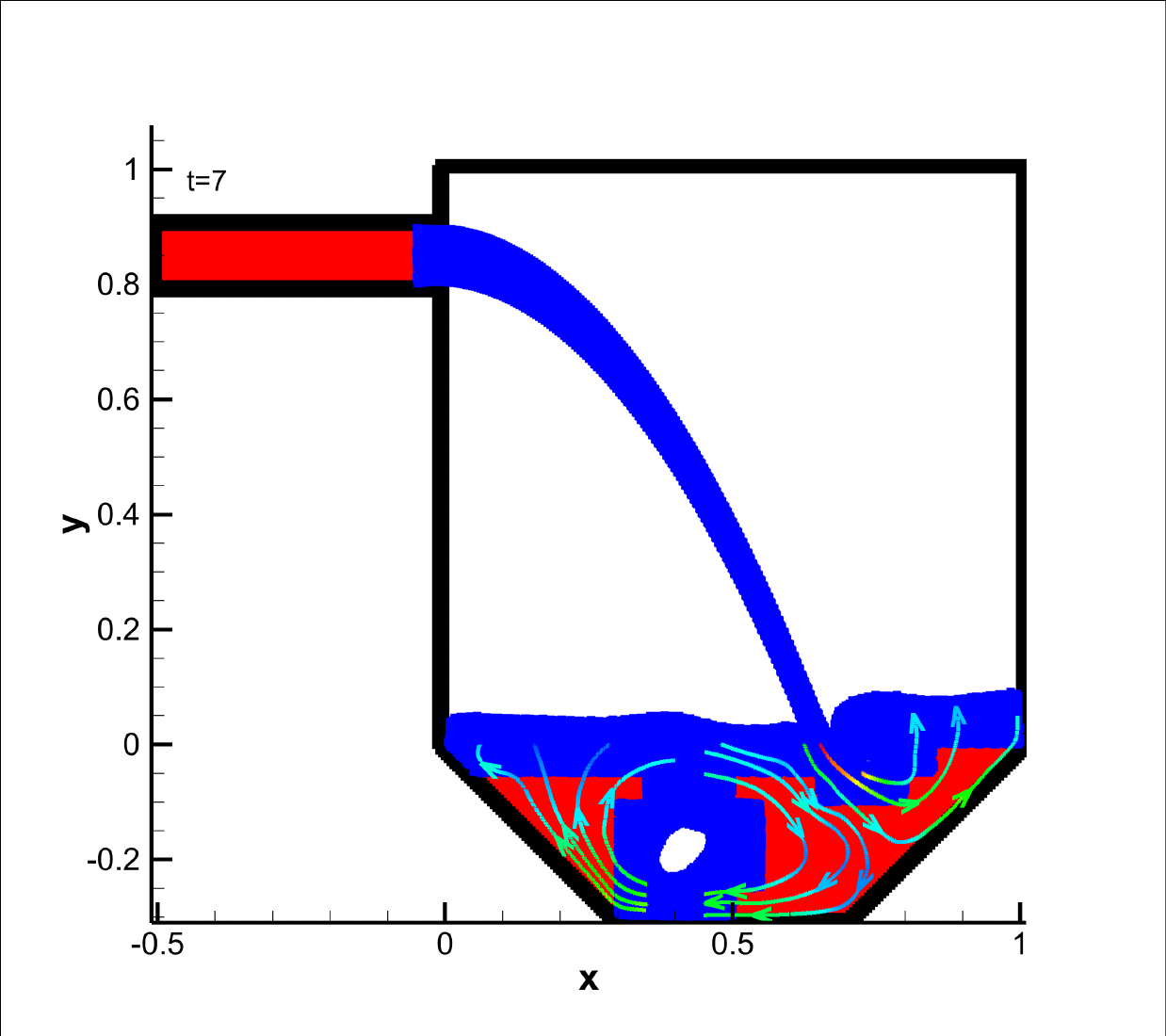}} \\
	\subfigure{\includegraphics[width = 0.45\columnwidth, trim = 10 10 10 10, clip]{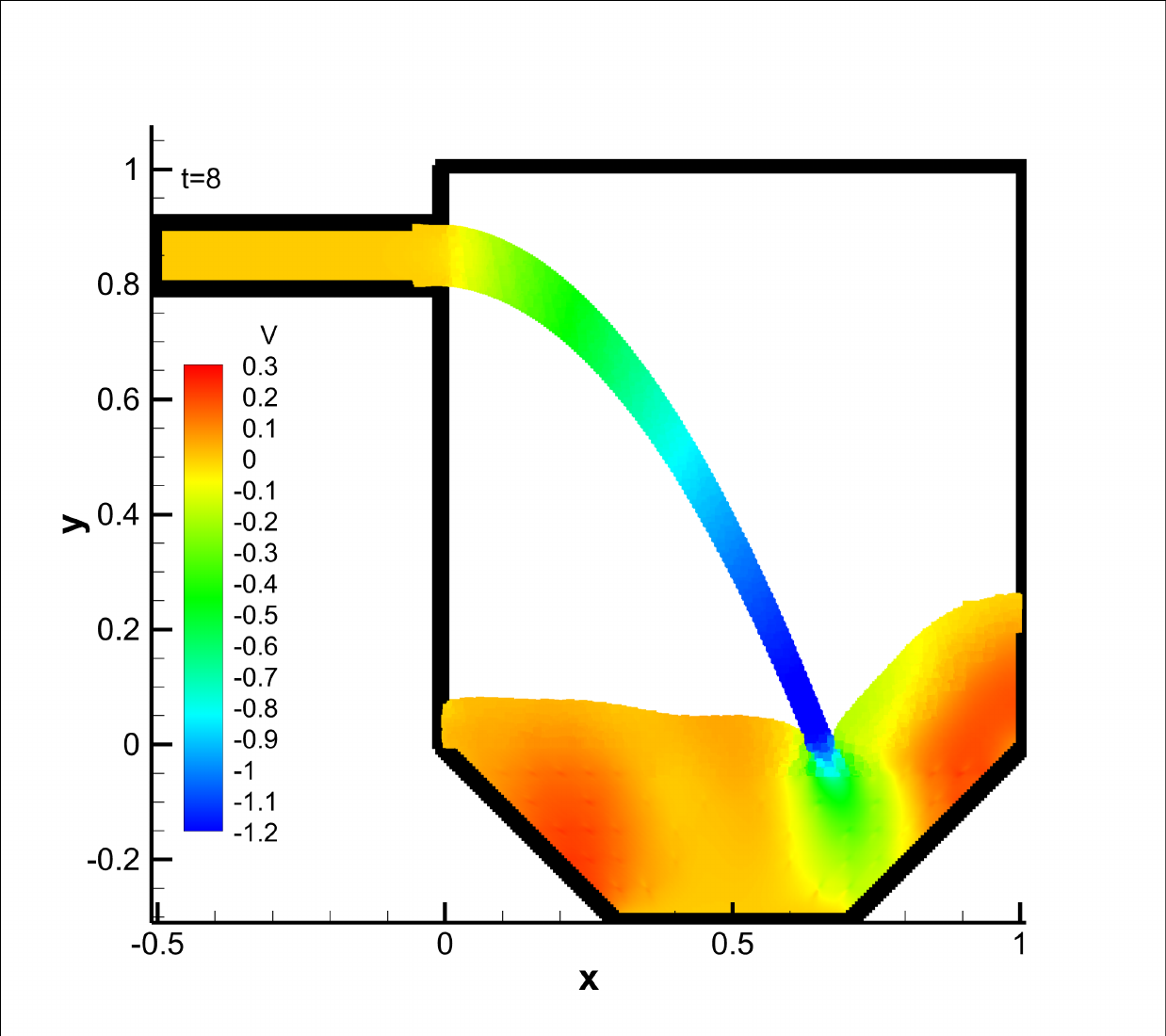}} \hspace{5pt}
	\subfigure{\includegraphics[width = 0.45\columnwidth, trim = 10 10 10 10, clip]{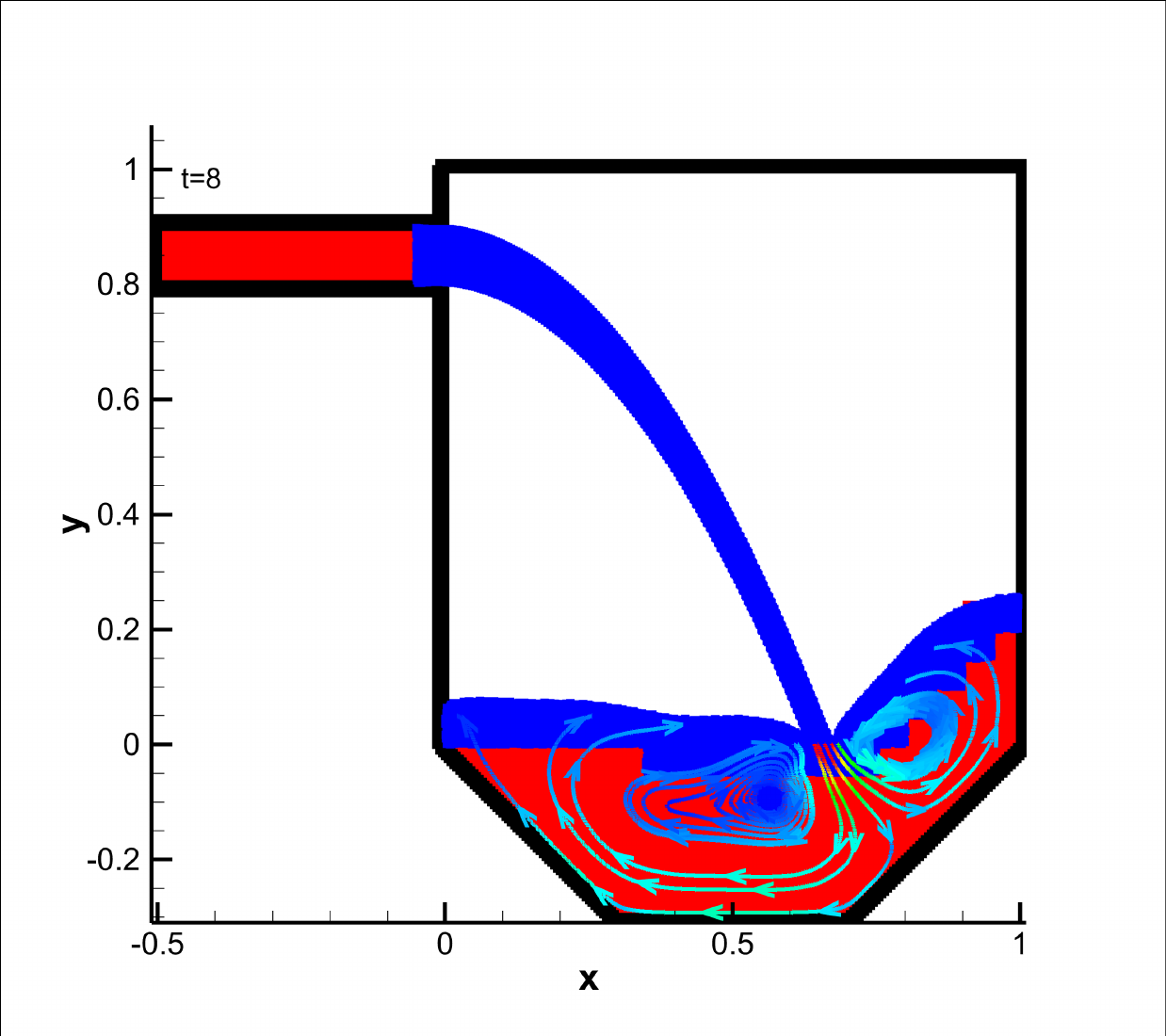}} \\
	\caption{Numerical results of water filling problem at $T = 6, 7, 8$. Left: $V$-component velocity, right: mesh-particle partitions (red: mesh region, blue: particle region) with streamlines}
	\label{fig:flooding2}
\end{figure}

\begin{figure}
	\centering
	\includegraphics[width=0.8\linewidth, trim = 10 10 10 10, clip]{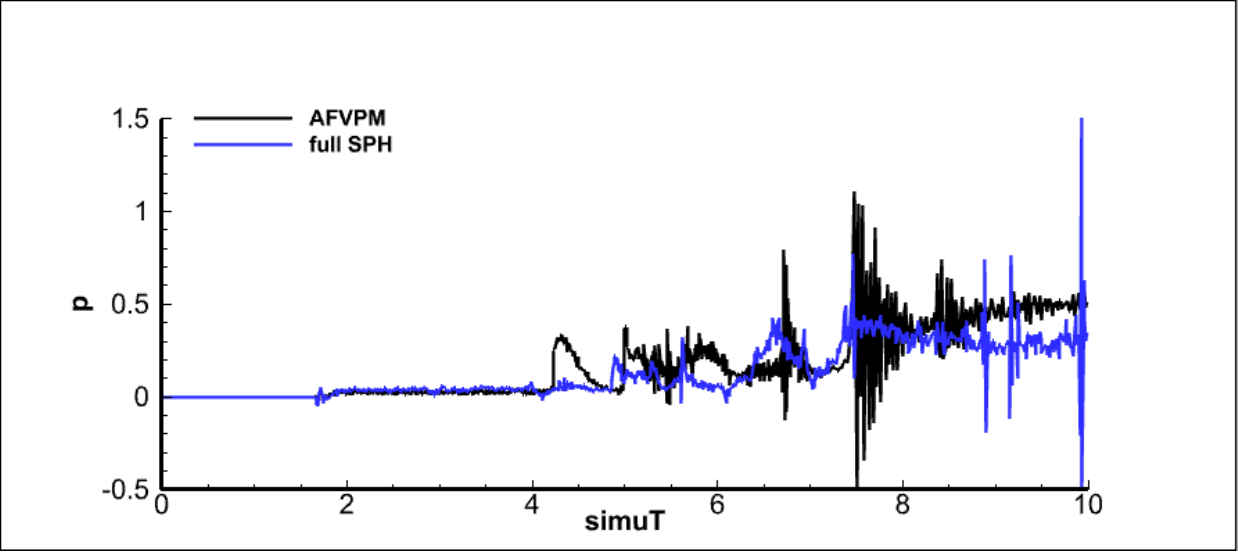}
	\caption{Pressure history at mid-point of bottom wall of water filling problem}
	\label{fig:fillingPress}
\end{figure}

The contours of $V$-component velocity and mesh-particle partitions at $T=2.5, 3.8, 4.5$ and $6, 7, 8$ are shown in Figure \ref{fig:flooding1} and \ref{fig:flooding2}, where water impacting, fragmentation and reconnection are clearly captured by AFVPM, and the mesh and particle regions converted to each other dynamically and adaptively with the movement of free surfaces, and the streamlines colored by magnitude of velocity in mesh regions are also shown in the figures of mesh-particle partitions. The pressure at the mid-point of bottom wall $(0.5, -0.3)$ is recorded every 500 iterations in Figure \ref{fig:fillingPress}, and AFVPM gives more flow details, e.g. the rolled water impacting at about $T = 4.2$.

\begin{figure}[!htbp]
	\centering
	\subfigure[$T=6$]{\includegraphics[width = 0.45\columnwidth, trim = 10 10 10 10, clip]{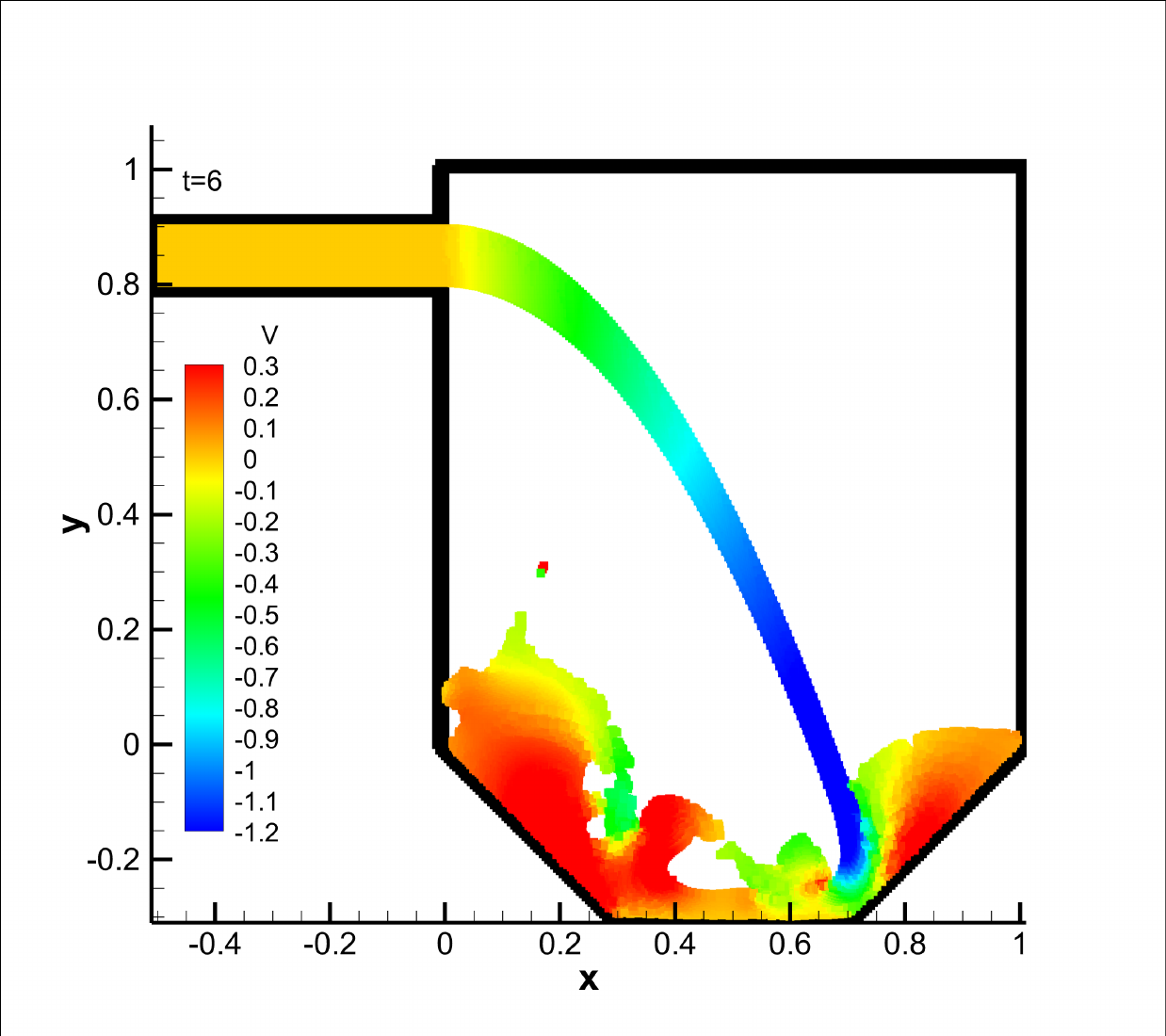}} \hspace{5pt}
	\subfigure[$T=7$]{\includegraphics[width = 0.45\columnwidth, trim = 10 10 10 10, clip]{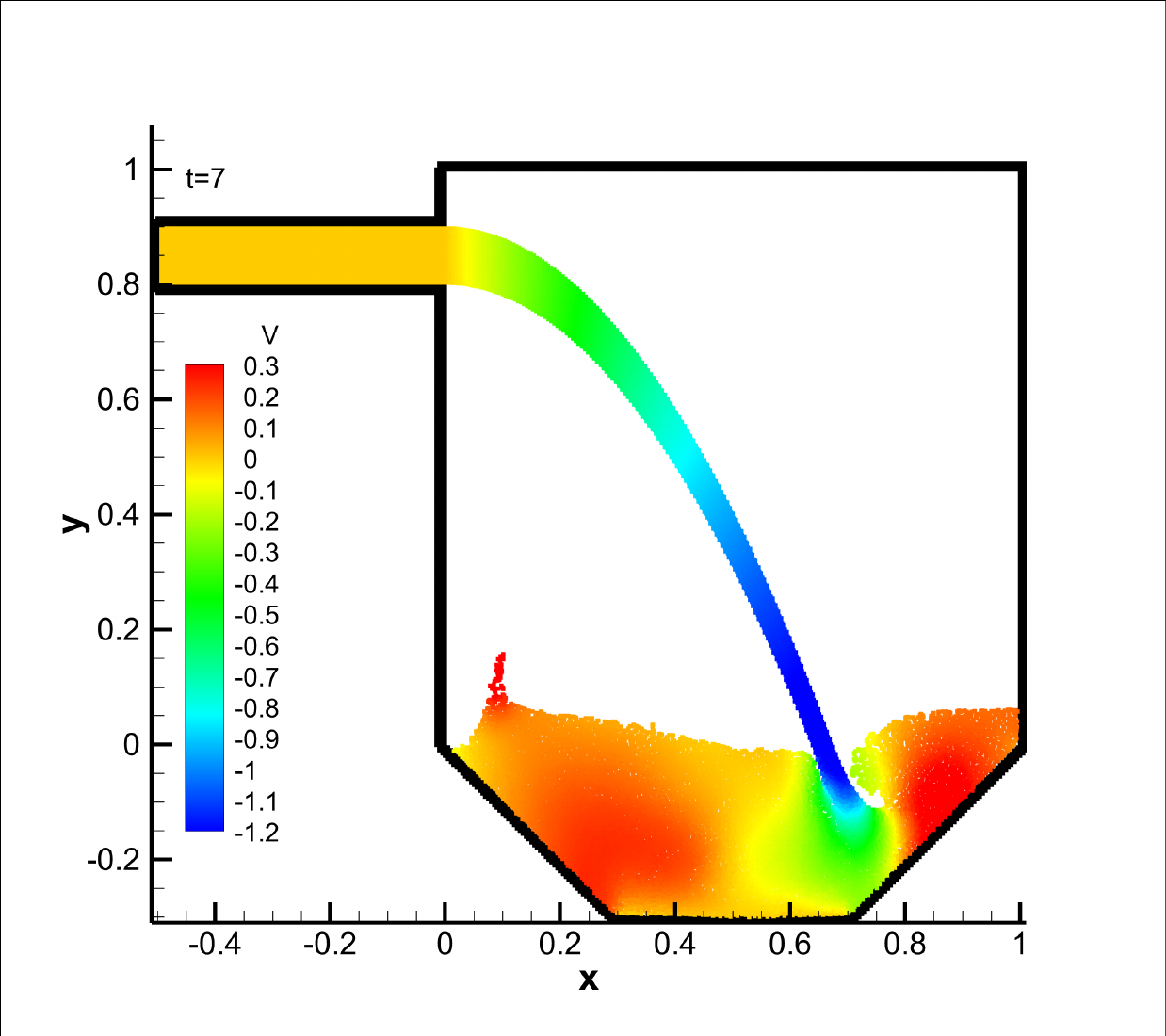}} \\
	\caption{$V$-component velocity of water filling problem at $T = 6, 7$ by full SPH approach}
	\label{fig:fullSPH-filling}
\end{figure}

This problem is also simulated by full SPH approach, and the contours of $V$-component velocity at $T = 6,7$ are shown in Figure \ref{fig:fullSPH-filling}. It is shown that the AFVPM has better resolution and accuracy for the free surface and the bubbles inside the water domain.

\subsection{Computational time analysis}
One of the main advantages of AFVPM with respect to full SPH is higher computational efficiency, since the majority of the computational domain is represented by mesh-based FVM. Table  \ref{tab:pureSPHGKS} shows that the CPU time needed by full SPH in still tank case is nearly three times of the full GKS method on identical particle and mesh cell numbers. Nevertheless, the coupled approach and treatment of mesh-particle interfaces need additional computational time.
\begin{table}[!htbp]
		\small
		\begin{center}
			\def\temptablewidth{1.0\textwidth}
			{\rule{\temptablewidth}{1pt}}
			\begin{tabular*}{\temptablewidth}{@{\extracolsep{\fill}}c c c}
				Method & CPU time / s\\
				\hline
				GKS & 21.9 \\ 	
                SPH & 62.5\\ 	
                Speed up ratio & 2.9
			\end{tabular*}
			{\rule{\temptablewidth}{1pt}}
		\end{center}
		\vspace{-4mm} \caption{ Computational efficiency of full GKS and full SPH}
		\label{tab:pureSPHGKS}
\end{table}

The computational time for the first 1000 steps of the AFVPM and full SPH method in the five numerical examples in Section \ref{sec:NumExp} are summarized in Table \ref{tab:efficiency}, where AFVPM can achieve about $150\%$ speedup than full SPH. 
\begin{table}[!htbp]
		\small
		\begin{center}
			\def\temptablewidth{1.0\textwidth}
			{\rule{\temptablewidth}{1pt}}
			\begin{tabular*}{\temptablewidth}{@{\extracolsep{\fill}}c c cc c c}
				Case & Initial $N_p + N_c$ & Mesh fraction (\%) & AFVPM & full SPH & speedup\\
				\hline
				Still tank & 20000 & 95 & 20.5 & 32.1 & 1.6 \\
                Dam breaking & 20000 & 81 & 32.3 & 45.6 & 1.41 \\
                Ship cruise & 54000 & 87 & 65.1 & 90.3 & 1.39 \\
                Body entry & 320000 & 95 & 459.2 & 711.7 & 1.55 \\
                Water filling (after $t > 8$) & 2000 & 72 & 36.0 & 46.4 & 1.29 \\
			\end{tabular*}
			{\rule{\temptablewidth}{1pt}}
		\end{center}
		\vspace{-4mm} \caption{ Computational efficiency of AFVPM with full SPH}
		\label{tab:efficiency}
\end{table}

\section{Conclusions}
This study proposes, for the first time, an adaptive finite volume-particle method (AFVPM) that enables dynamic and adaptive two-way conversion between Eulerian meshes and Lagrangian particles in response to the evolution of moving free interfaces. AFVPM effectively combines the complementary strengths of Eulerian finite volume and Lagrangian SPH methods for free-surface flow simulation. By strategically employing FVM in bulk-flow regions and SPH in the vicinity of free surfaces, AFVPM achieves significant improvements in both computational efficiency and accuracy. The proposed adaptive two-way conversion strategy, together with a robust mesh-particle interface algorithm, ensures seamless coupling between the two discretization frameworks. Comprehensive validation using standard benchmark cases confirms that AFVPM delivers high accuracy in free-surface flow simulations while achieving about $150\%$ speedup over full SPH methods. Future work will focus on extending the framework to multiphase flows and parallelizing the algorithm for large-scale simulations. Overall, the present methodology provides a solid foundation for the development of next-generation hybrid solvers that balance accuracy, robustness, and computational efficiency in interfacial flow modeling.

\section*{Acknowledgements}
We are grateful to Dr. Xiaojian Yang and Dr. Yaqing Yang for their helpful discussions on numerical analysis, assistance with figure plotting, and valuable comments on the manuscript.
This research was supported by the Research Grants Council Areas of Excellence (AoE) Scheme (AoE/P-601/23N-D- MATH), and by CORE as a joint research center for ocean
research between Laoshan Laboratory and HKUST.

\bibliographystyle{ieeetr}
\bibliography{reference}

\end{document}